\documentclass[twocolumn,amssymb,fleqn]{revtex4} %fleqn: make all eqns left alighed  ,showpacs
\usepackage{epsfig,amssymb,amsmath,graphicx,subfigure,hyperref}  % pdfpages and graphicx are added   hyperref(set hyperlink for content)
\usepackage{color}

\usepackage{multirow} 
\usepackage{tabularx}

\begin{document}

\title{Ordering leads to multiple fast tracks in simulated collective escape of human crowds}
\author{Chen Cheng,\textit{$^{a}$} Jinglai Li,\textit{$^{b}$} and Zhenwei Yao\textit{$^{c}$}}
\email{zyao@sjtu.edu.cn}
\affiliation{$^{a}$~School of Mathematical Sciences, Shanghai Jiao Tong University, Shanghai 200240, China\\
$^{b}$~School of Mathematics, University of Birmingham, Edgbaston, Birmingham B15 2TT, UK\\
$^{c}$~School of Physics and Astronomy, and Institute of Natural Sciences, Shanghai Jiao Tong University, Shanghai 200240, China}

\begin{abstract}
Elucidating emergent
  regularities in intriguing crowd dynamics is a fundamental scientific problem
  arising in multiple fields. In this work, based on the social force model, we
  simulate the typical scenario of collective escape towards a single exit and
  reveal the striking analogy of crowd dynamics and crystallisation.  With the
  outflow of the pedestrians, crystalline order emerges in the compact crowd.
  In this process, the local misalignment and global rearrangement of
  pedestrians are well rationalized in terms of the characteristic motions of
  topological defects in the crystal. Exploiting the notions from the physics of
  crystallisation further reveals the emergence of multiple fast tracks in the
  collective escape.
\end{abstract}

\maketitle

\section{Introduction}

Collective motion of pedestrians is a common phenomenon in various
daily scenarios of urban life, and once driven by life-threatening panic,
overcrowding and stampede may lead to fatalities~\cite{
Helbing1995Social,helbing2005self,helbing2007dynamics, gravish2015glass, tao2017floor,rogsch2010panic,
adrian2019glossary,gibelli2019crowd,rahouti2018evacuation,feliciani2018investigation}.
Understanding crowd dynamics is thus crucial for achieving safety and efficiency
at both individual and collective levels~\cite{smith1993engineering,
Helbing2000VICSEK, batty2003discrete, metivet2018how}. Observations show that the 
majority of pedestrians walk in groups~\cite{moussaid2010walking}, and the size of 
free-forming groups conforms to Poisson distribution, indicating the existence
of statistical regularity in complicated crowd behaviours~\cite{james1953distribution,
sumpter2006principles,Castellano2009Statistical}.  Collective phenomena,
such as oscillational behaviour~\cite{schadschneider2009evacuation}
and the fast-is-slower~\cite{lakoba2005modifications,zuriguel2020contact} and
stop-and-go~\cite{chraibi2015jamming,boltes2018empirical,cordes2019trouble}
effects, are experimentally observed
~\cite{Helbing2000VICSEK,helbing2009pedestrian}. Quantitative
measurement of pedestrian flows further reveals an analogy with the
Navier-Stokes equations that originally describe dynamics of
fluids~\cite{henderson1971statistics,henderson1974fluid}. The phenomenological
fluid-dynamic approach captures macroscopic behaviours of crowd dynamics by
averaging the behaviours of neighbouring individuals~\cite{helbing1998fluid,
bauer2007macroscopic}.

Further scrutiny of individual pedestrians could provide valuable microscopic
information that yields insights into the intriguing crowd
dynamics~\cite{Helbing2000VICSEK,helbing2003lattice,Castellano2009Statistical,Czir2012Collective,chraibi2018modelling,corbetta2018physics,das2020introduction}.
Following this idea, a social force model~{(SFM)} has been proposed to simulate the motion of human
crowds~\cite{helbing1991mathematical, Helbing1995Social,Helbing2000VICSEK}. 
The basic idea underlying the SFM is to treat pedestrians 
as particles with a simplified will. Specifically, this model features a mixture of both physical and
socio-psychological forces influencing the walking behaviour of pedestrians, and
it has the unique advantage of incorporating new forces as our understanding of
pedestrians is
advanced~\cite{Helbing1995Social,Helbing2000VICSEK,moussaid2010walking}. 
Much has been learned about the law of crowd dynamics by applying the SFM in various
situations~\cite{Helbing2000VICSEK,helbing2007dynamics,moussaid2010walking,Colombi2017Modelling}.

While the SFM could conveniently provide detailed dynamical
information of crowd motion by specifying proper values of model
parameters~\cite{Helbing2000VICSEK}, the challenge is to exploit the simulated
data from a suitable perspective for revealing the underlying regularities.  
In this work, we resort to the analogy of crowd dynamics and crystallisation
process~\cite{nelson2002defects,yao2014polydispersity,yao2016dressed},
and gain insights into the intriguing collective motion.  Specifically, we
simulate the typical scenario of collective escape towards a single exit using
the generalized social force model  that incorporates the random behaviours of
pedestrians. Our simulations show the rapid ordering of the initially randomly
distributed pedestrians. In the compact ``crystallised'' crowd, local
misalignments emerge at random sites, which are recognised as topological
defects in two-dimensional crystal~\cite{chaikin2000principles,
nelson2002defects}. With the outflow of pedestrians, the microscopic crystalline
structure underlying the crowd is under persistent transformation, which exactly
corresponds to the characteristic annihilation and glide motion of topological
defects in the crystal. From the striking analogy between crowd dynamics and
crystallisation process, and in combination with statistical analysis, we
demonstrate the emergence of multiple fast tracks resulting from the
spontaneously formed crystal structure in the escaping crowd. This work reveals
the regularities in crowd dynamics from the perspective of crystallisation, and
may provide useful information for understanding crowd behaviours in
evacuation.

\section{Model and method}

We simulate the crowd dynamics of pedestrians based on the SFM~\cite{Helbing2000VICSEK}. 
It assumes a mixture of physical and socio-psychological forces influencing the crowd behaviour 
by considering personal motivations and environmental
constraints. In this model, each pedestrian $i$ of mass $m_i$ and
velocity $\boldsymbol{v}_i$ tends to move by a
desired speed $v_i^p$ along a certain direction $\boldsymbol{e}_i^p$ during the acceleration
time $\tau_i$. The resulting personal desire force $\boldsymbol{F}_p$ is: 
\begin{eqnarray} \label{eq:1} 
\boldsymbol{F}_p = m_i\frac{v_i^p\boldsymbol{e}_i^p-\boldsymbol{v}_i}{\tau_i}. 
\end{eqnarray}
Here, we note that the pedestrians are polarised to proceed towards
the exit, since the desired direction of motion always points to
the exit in the model. Furthermore, 
pedestrians psychologically tend to keep a social distance between each other 
and avoid hitting walls. This is modelled by introducing ``interaction force'' 
$\boldsymbol{f}_{ij}$ between pedestrians $i$ and $j$ and $\boldsymbol{f}_{iW}$ between 
pedestrian $i$ and the wall, respectively. The total interaction force is
\begin{eqnarray}\label{eq:2} 
\boldsymbol{F}_{int} = \sum\limits_{j(\neq i)}\boldsymbol{f}_{ij} + \sum\limits_{W}\boldsymbol{f}_{iW}. 
\end{eqnarray} 
Combining eqn~(\ref{eq:1}) and~(\ref{eq:2}), we obtain the acceleration equation
\begin{eqnarray} \label{eq:3} 
m_i\frac{\rm{d} \boldsymbol{v}_i}{\rm{d} t} = m_i\frac{v_i^p(t)\boldsymbol{e}_i^p(t) 
- \boldsymbol{v}_i(t)}{\tau_i} 
+ \sum\limits_{j(\neq i)}\boldsymbol{f}_{ij} 
+ \sum\limits_{W}\boldsymbol{f}_{iW}. 
\end{eqnarray} 
The position vector $\boldsymbol{r}_i(t)$ is updated by the velocity $\boldsymbol{v}_i(t)
  = \rm{d} \boldsymbol{r}_i/\rm{d} t$.

 The interaction force $\boldsymbol{f}_{ij}$ between pedestrian $i$ and $j$ is
 specified as follows.  With the distance $d_{ij} = ||\boldsymbol{r}_i - \boldsymbol{r}_j||$
 between the two pedestrians' centres of mass, the psychological tendency of
 pedestrian $i$ to stay away from pedestrian $j$ is described by a repulsive
 interaction force $A_i{\rm exp}[(r_{ij} - d_{ij})/B_i]\boldsymbol{n}_{ij}$, where
 $A_i$ and $B_i$ are constants, indicating the strength and the range of the
 interaction, and $\boldsymbol{n}_{ij} = (n_{ij}^1, n_{ij}^2) = (\boldsymbol{r}_i -
 \boldsymbol{r}_j)/d_{ij}$ is the normalised directional vector pointing from
 pedestrian $j$ to $i$. The pedestrians touch each other if their distance
 $d_{ij}$ is smaller than the sum $r_{ij} = r_i + r_j$ of their radius $r_i$ and
 $r_j$. In our model, we specify a uniform value for the size of each
 pedestrian (see Table~\ref{tab.1})~\cite{Helbing2000VICSEK}. While in reality the size of human body 
 among the crowd is not uniform, previous work on 2D crystallisation shows that the 
 size-polydispersity effect must be strong enough to disrupt the crystalline order~\cite{yao2014polydispersity}. 
 The factor of size polydispersity among adult pedestrians is thus ignored in this work. 
 Inspired by granular interactions, two additional forces are included in the
 model, which are essential for understanding the particular effects in
 panicking crowds: a ``body force'' $k(r_{ij} - d_{ij})\boldsymbol{n}_{ij}$
 counteracting body compression and a ``sliding friction force'' $\kappa(r_{ij}
 - d_{ij})\Delta v_{ji}^t\boldsymbol{t}_{ij}$ impeding relative tangential motion, if
 pedestrians $i$ and $j$ are close enough. Here $\boldsymbol{t}_{ij} = (-n_{ij}^2, n_{ij}^1)$
 means the tangential direction and $\Delta v_{ji}^t = (\boldsymbol{v}_j -
 \boldsymbol{v}_i)\cdot\boldsymbol{t}_{ij}$ the tangential velocity difference, while $k$
 and $\kappa$ are large constants, representing the bump and the friction
 effect. In summary, the interaction force $\boldsymbol{f}_{ij}$ between pedestrians
 $i$ and $j$ is given by
 \begin{equation}\label{eq:4} 
 \begin{split}
\boldsymbol{f}_{ij} =& \{A_i{\rm exp}[(r_{ij} - d_{ij})/B_i] + kg(r_{ij} - d_{ij})\}\boldsymbol{n}_{ij}\\ 
&+ \kappa g(r_{ij} - d_{ij})\Delta v_{ji}^t\boldsymbol{t}_{ij},
\end{split}
\end{equation}
where the indicator function $g(r_{ij} - d_{ij})$ is zero for
$r_{ij} - d_{ij} < 0$  and it is equal to $r_{ij} - d_{ij}$ otherwise.

The interaction with the walls is treated analogously. By denoting $d_{iW}$ as the distance 
to wall $W$, $\boldsymbol{n}_{iW}$ as the direction perpendicular to it, and $\boldsymbol{t}_{iW}$ as the 
direction tangential to it, we have
\begin{equation}\label{eq:5} 
\begin{split}
\boldsymbol{f}_{iW} =& \{A_i{\rm exp}[(r_i - d_{iW})/B_i] + kg(r_i - d_{iW})\}\boldsymbol{n}_{iW}\\ 
&- \kappa g(r_i - d_{iW})(\boldsymbol{v}_i\cdot\boldsymbol{t}_{iW})\boldsymbol{t}_{iW}.
\end{split}
\end{equation}

\begin{figure}[th]
\centering
\includegraphics[width=.45\textwidth]{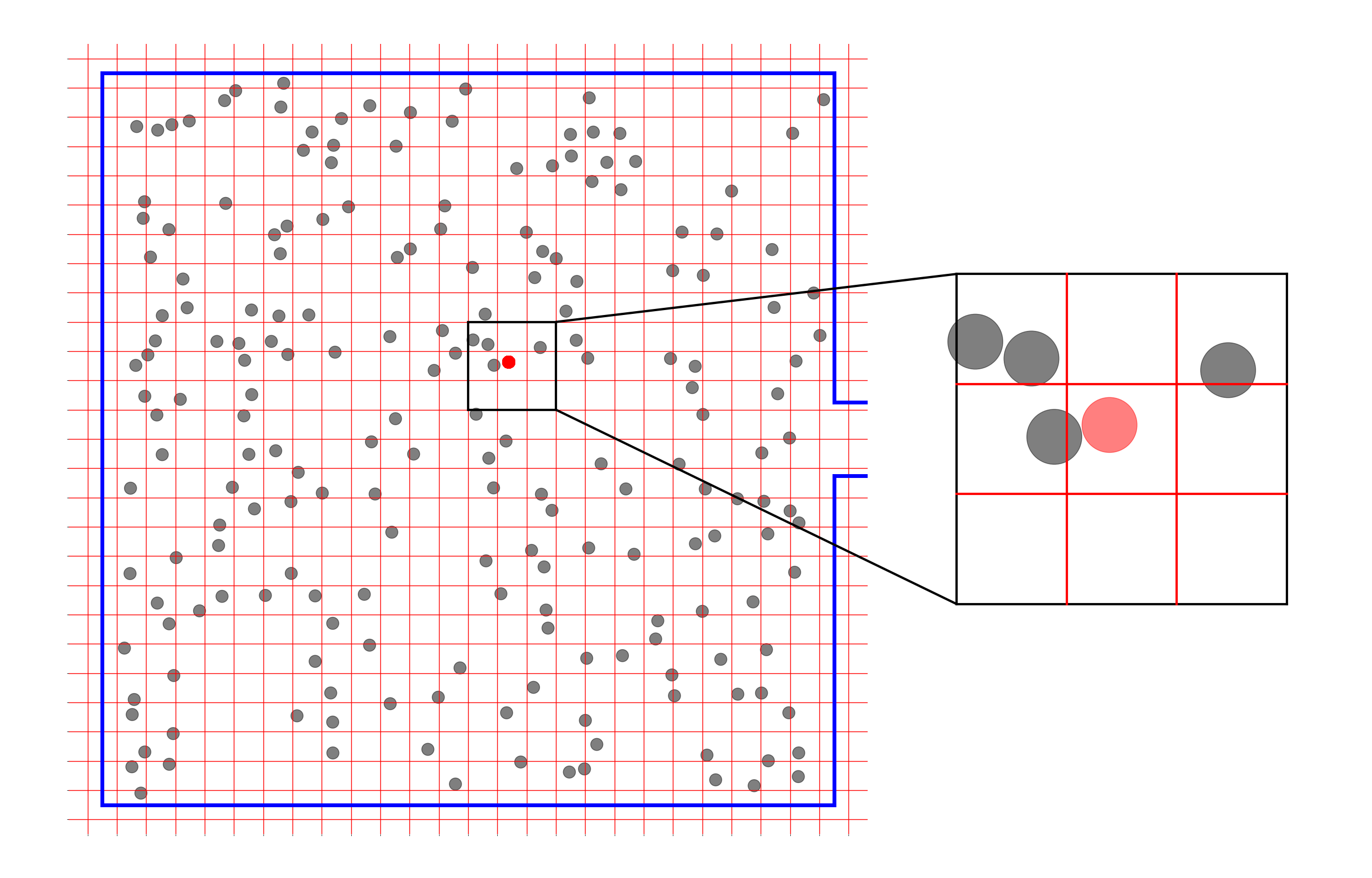}
\caption{Schematic plot 
  of the simulation scenario. Pedestrians are depicted by dots and walls are depicted by blue 
  lines with an open door in the right wall. Red grids are introduced to implement the 
  cell-list method for efficiently computing the interaction forces. See the main text for more information.}
\label{fig.1}
\end{figure}

In the implementation of the SFM for large-scale systems, the most
time-consuming part is to compute the interaction forces. For a $N$-particle
system, the maximum number of pairwise interactions is $N(N-1)/2$, which leads
to $O(N^2)$ time complexity. In this work we adopt the cell-list
method~\cite{matin2003cell} to avoid repeated calculation of pairwise
interactions and thus significantly reduce the time complexity from $O(N^2)$ to
$O(N)$~\cite{dobson2016cell}. The cell-list method achieves this goal by
dividing the physical space of the system into equal grids called cells. Each
particle is assigned to a specific cell. In a two-dimensional system, each cell
has eight neighbouring cells and a cell together with its eight neighbouring
cells are called a cell neighbourhood (see Fig.~\ref{fig.1}). For each
particle, we only need to calculate the interacting forces between the particles
within the cell neighbourhood.

\section{Results and discussion}

In this work we use the SFM to study a typical scenario of collective escape
towards a single exit (see Fig.~\ref{fig.1}). Initially, $N$ pedestrians are
randomly distributed in a square room of side length $H_0$, and the
parameter values used in our simulation are presented in Table~\ref{tab.1},
which largely follow Ref.12.  For simplicity, the values
for $m_i$, $v^p_i$, $\tau_i$, $r_i$, $A_i$ and $B_i$ for each pedestrian are
taken to be identical. The moment when the crowd start to run towards the exit
is denoted as $t=0$. In our simulation, the state of motion is updated according
to eqn~(\ref{eq:3}) by the time step $\Delta t$. In particular, the interaction
forces between pedestrians are computed by the cell-list method. The edge
length of each cell is set to be $2r + 5B$ to ensure that the contribution from
the next nearest neighbouring cells is negligibly small. The escape time for
individual pedestrians is denoted as $T_{es}$, and the evacuation time for all
the pedestrians to leave the room is denoted as $T_{ev}$.

\begin{table}[h]
\small
\caption{List of parameters.}
\label{tab.1}
\begin{tabular*}{0.48\textwidth}{@{\extracolsep{\fill}}lll}
 \hline
 Variable&Value&Description \\
 \hline
  $H_0$ & $30\,\rm{m}$ & side length of the room\\
  $N$ & $1000$ & number of pedestrians\\
  $m$ &  $80\,\rm{kg}$  & mass of pedestrians \\
  $v^p$ & $1.0\,\rm{m/s}$   & desired velocity \\
  $\tau$ & $0.5\,\rm{s}$  & acceleration time \\
  $r$ & $0.3\,\rm{m}$  & radius of pedestrians\\
  $A$ & $2\times 10^3\,\rm{N}$  & interaction strength\\
  $B$ & $0.08\,\rm{m}$ & interaction range\\
  $k$ & $1.2\times 10^5\,\rm{kg/s^2}$ & bump effect \\
  $\kappa$ & $2.4\times 10^5\,\rm{kg/(m\cdot s)}$ & friction effect \\
  $\Delta t$ & $0.001\,\rm{s}$ & time step in simulation\\
 \hline
\end{tabular*}
\end{table}

\subsection{Collective escape statistics}

\begin{figure}[t] 
  \centering
  \subfigure[]{\includegraphics[width=.39\textwidth]{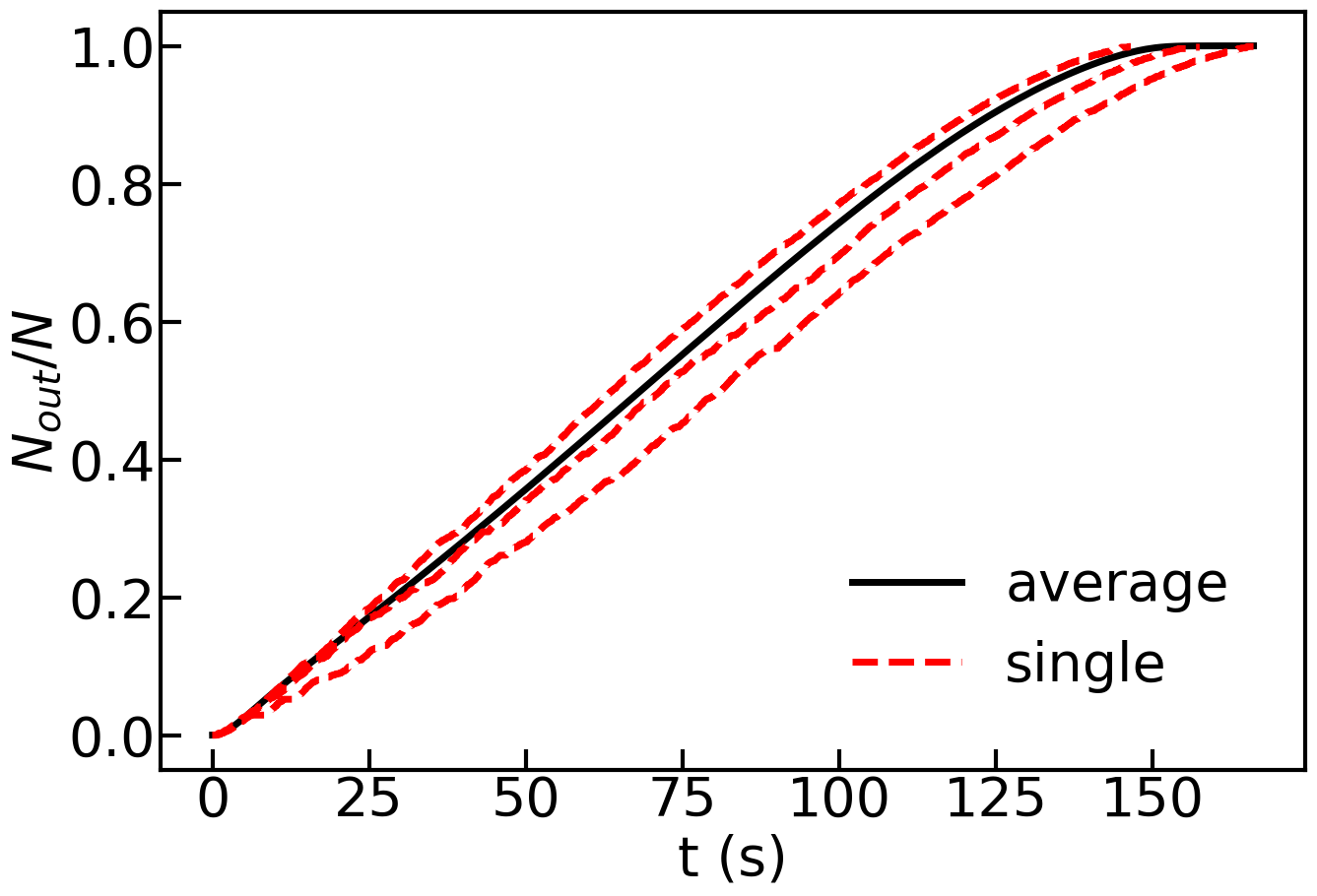}\label{fig.survival}}
  \subfigure[]{\includegraphics[width=.39\textwidth]{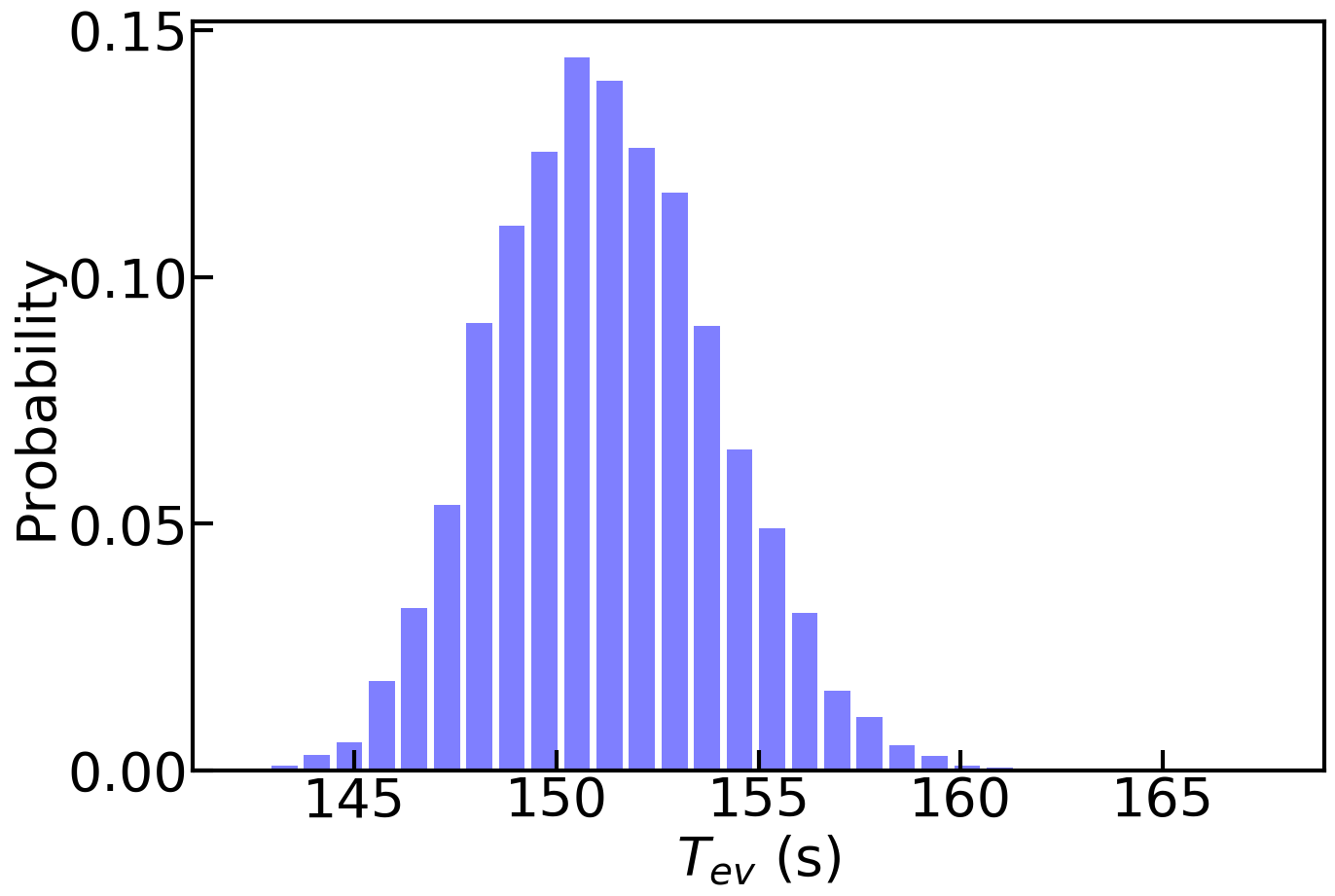}\label{fig.hist}}
  \caption{Statistical analysis of the instantaneous escape rate
  in a typical scenario of collective escape towards a single exit. The escape rate is defined as the
  ratio $N_{out}/N$, where $N_{out}$ is the number of pedestrians leaving the
  room. (a) Plot of $N_{out}/N$ versus time. The black solid curve is the
  average instantaneous escape rate over 6000 independent simulations with
  random initial conditions. The escape rates of randomly picked simulation
  runs are also plotted in red dashed curves. (b) The distribution of the
  evacuation time $T_{ev}$ for the 6000 simulations.}
  \label{fig:2}
\end{figure}

We analyze several statistics regarding escape of the pedestrians. 
First, the simulations allow us to track the instantaneous collective escape rate, 
which is defined as the percentage of the pedestrians who have successfully exited the room: 
$N_{out}/N$ where $N_{out}$ is the number of escaped pedestrians. The rate $N_{out}/N$ 
is plotted against time $t$ in Fig.~\ref{fig.survival}. 
The black curve is the average instantaneous escape rate over 6000 independent
simulations with random initial conditions, and the red dashed curves show the
results of several randomly selected sample simulation trials. 
From the figure, we see that, despite of the highly nonlinear
behaviors of individual pedestrians, the collective escape rate exhibits a
rather linear dependence on time over a long duration, indicating a steady
outflow of pedestrians. The escape curve becomes saturated only at the end of
the escape event, and the plateau lasts for a relatively short duration. 
The evacuation time $T_{ev}$ of independent simulations is also analyzed
statistically, and the results are shown in Fig.~\ref{fig.hist}. The histogram
of $T_{ev}$ is approximately a normal distribution. The curve of the collective
escape rate $N_{out}/N$ and the histogram of the evacuation time $T_{ev}$ imply
the existence of statistical regularities underlying the highly complicated
individual motions of pedestrians.

\subsection{Formation of multiple fast tracks}

\begin{figure*}[th]
\hspace{-6mm}
\subfigure[]{
  \begin{minipage}{0.25\textwidth}
    \centering
    \includegraphics[width=\linewidth]{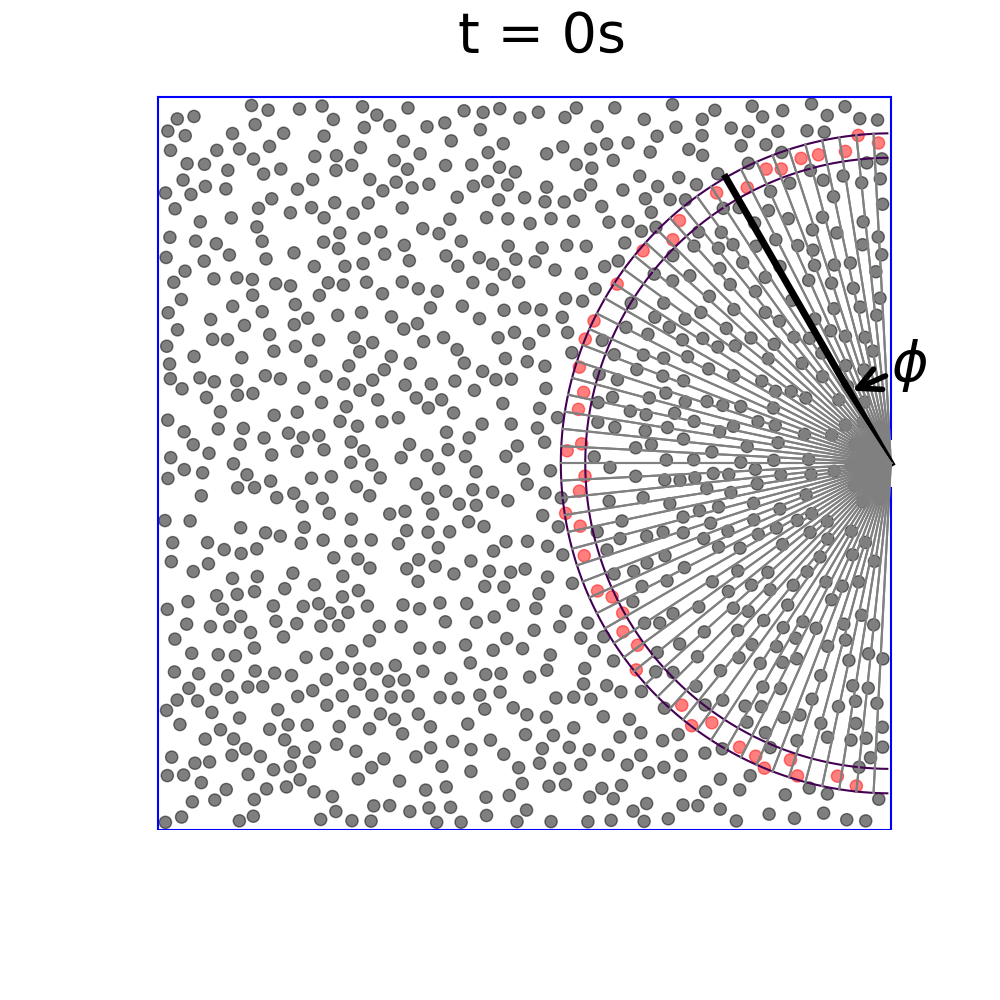}\\
    \vspace{-5mm}
    \includegraphics[width=\linewidth]{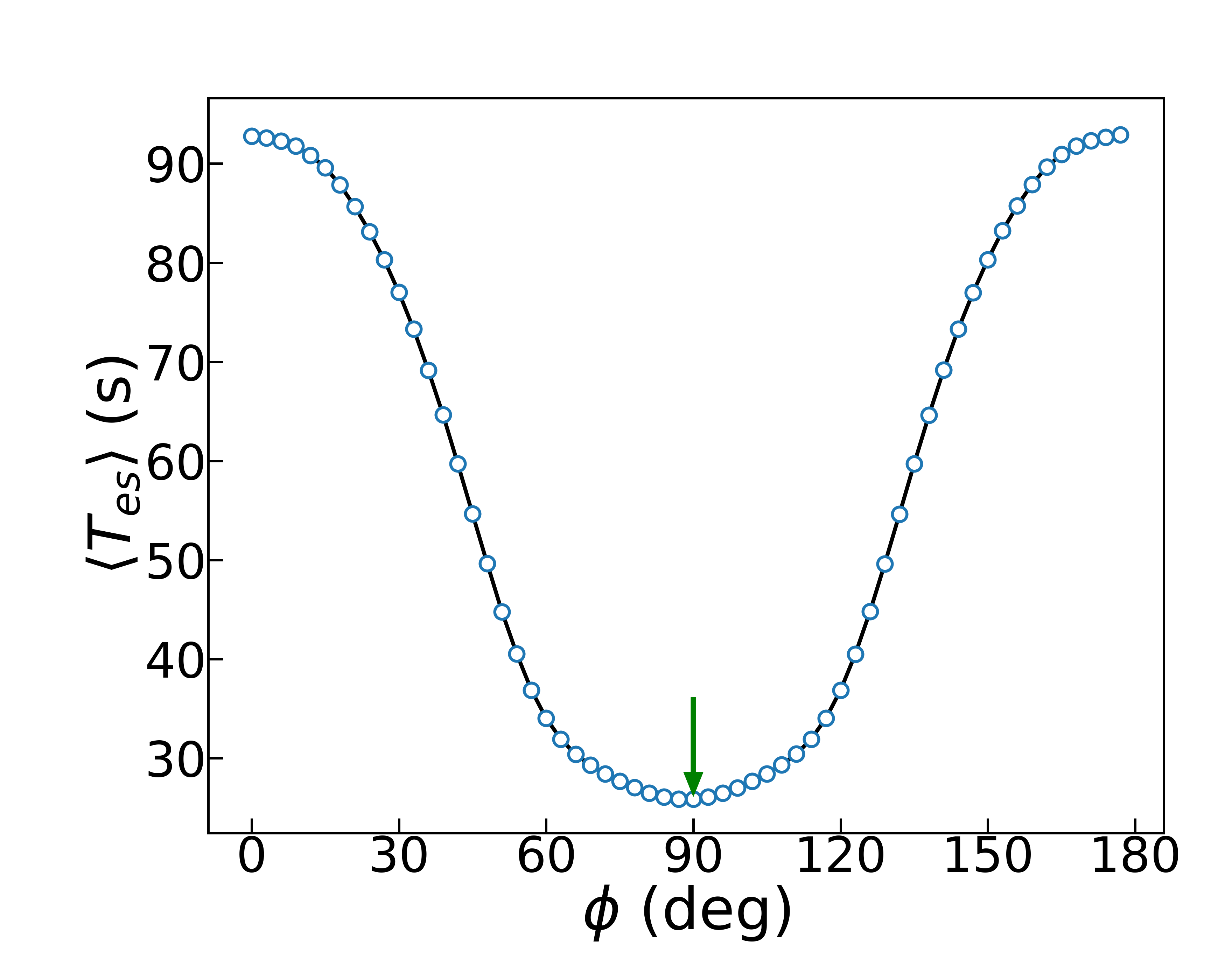}\\
  \end{minipage}
  \label{fig:3a}
}
\hspace{-6mm}
\subfigure[]{
  \begin{minipage}{0.25\textwidth}
    \centering
    \includegraphics[width=\linewidth]{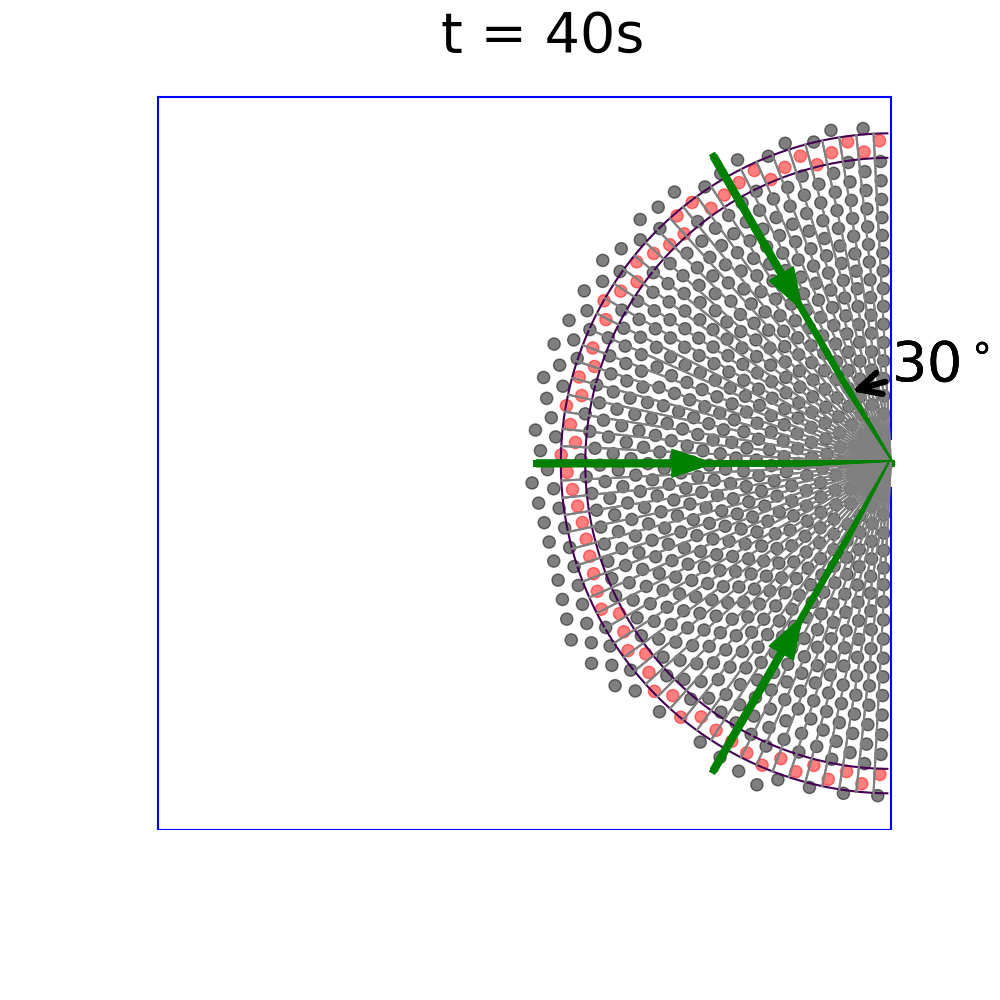}\\
    \vspace{-5mm}
    \includegraphics[width=\linewidth]{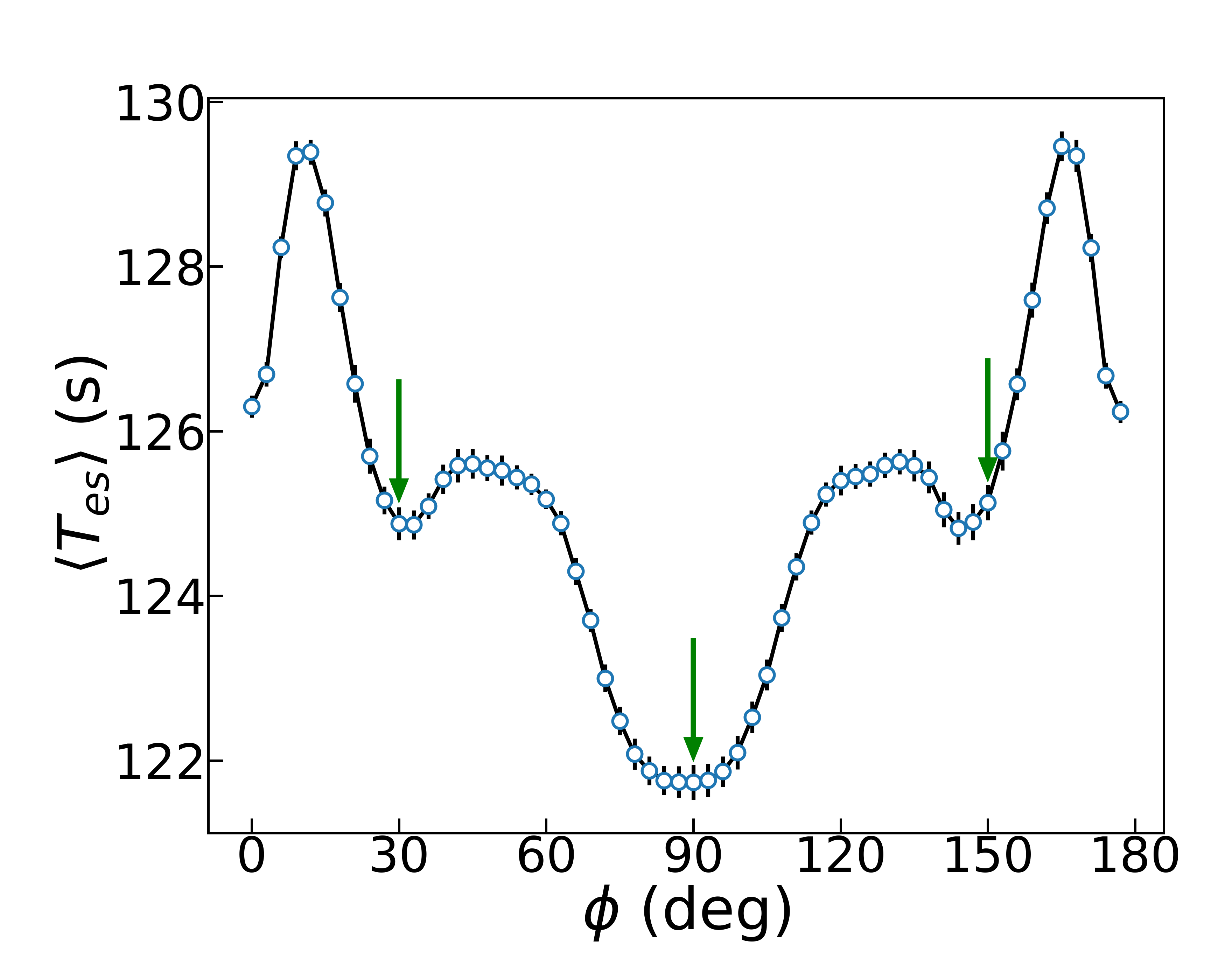}\\
  \end{minipage}
  \label{fig:3b}
}
\hspace{-4mm}
\subfigure[]{
  \begin{minipage}{0.25\textwidth}
    \centering
    \includegraphics[width=\linewidth]{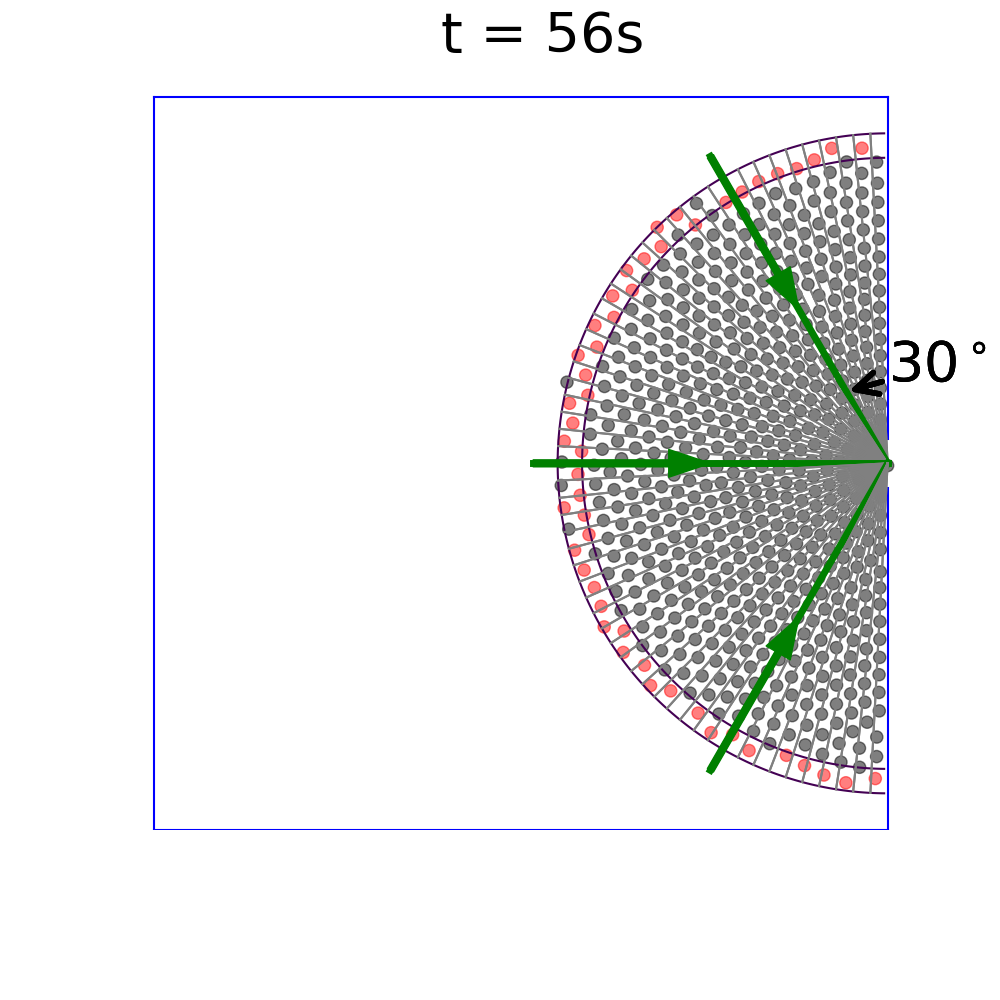}\\
    \vspace{-5mm}
    \includegraphics[width=\linewidth]{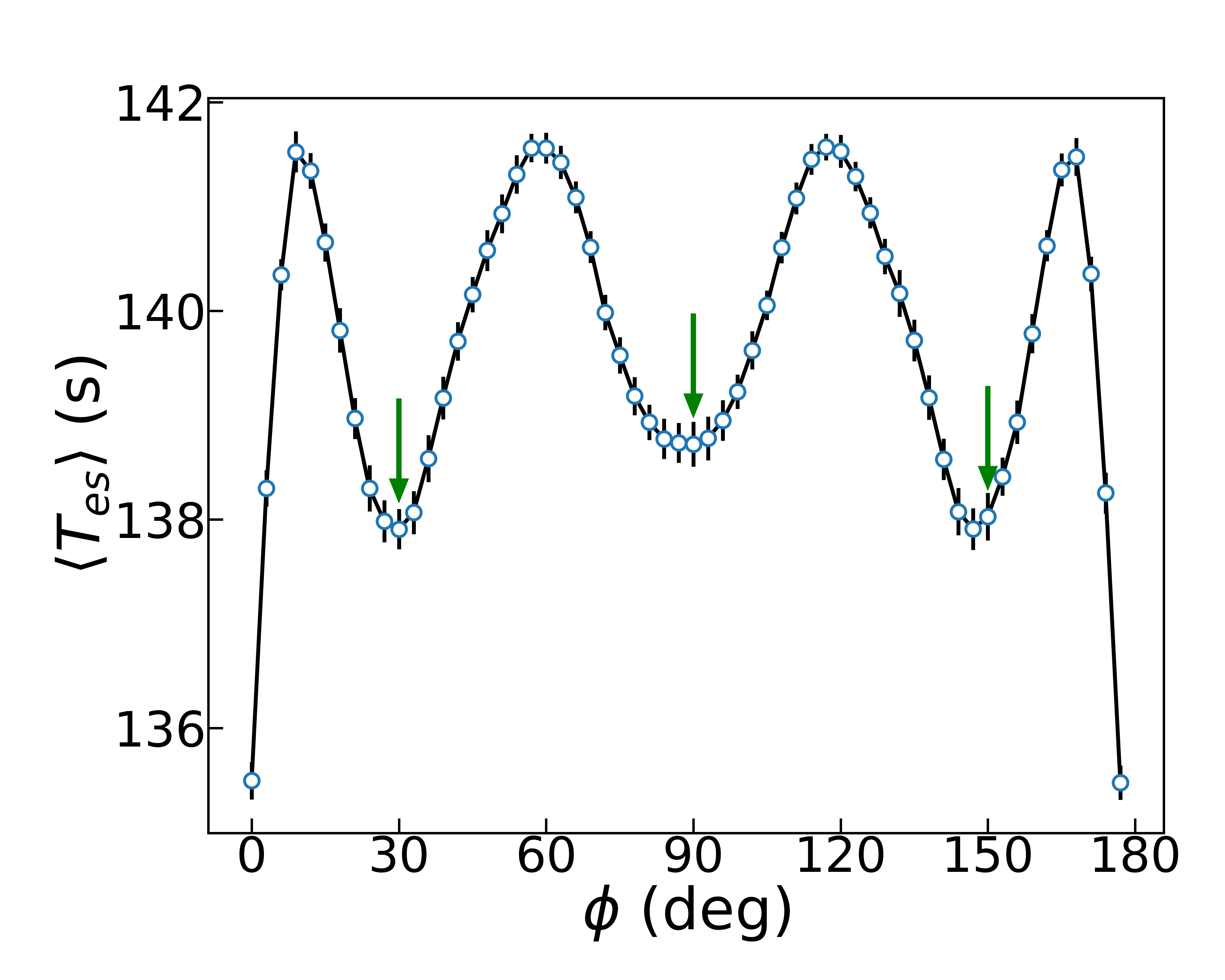}\\
  \end{minipage}
  \label{fig:3c}
}
\hspace{-5mm}
\subfigure[]{
  \begin{minipage}{0.25\textwidth}
    \centering
    \includegraphics[width=\linewidth]{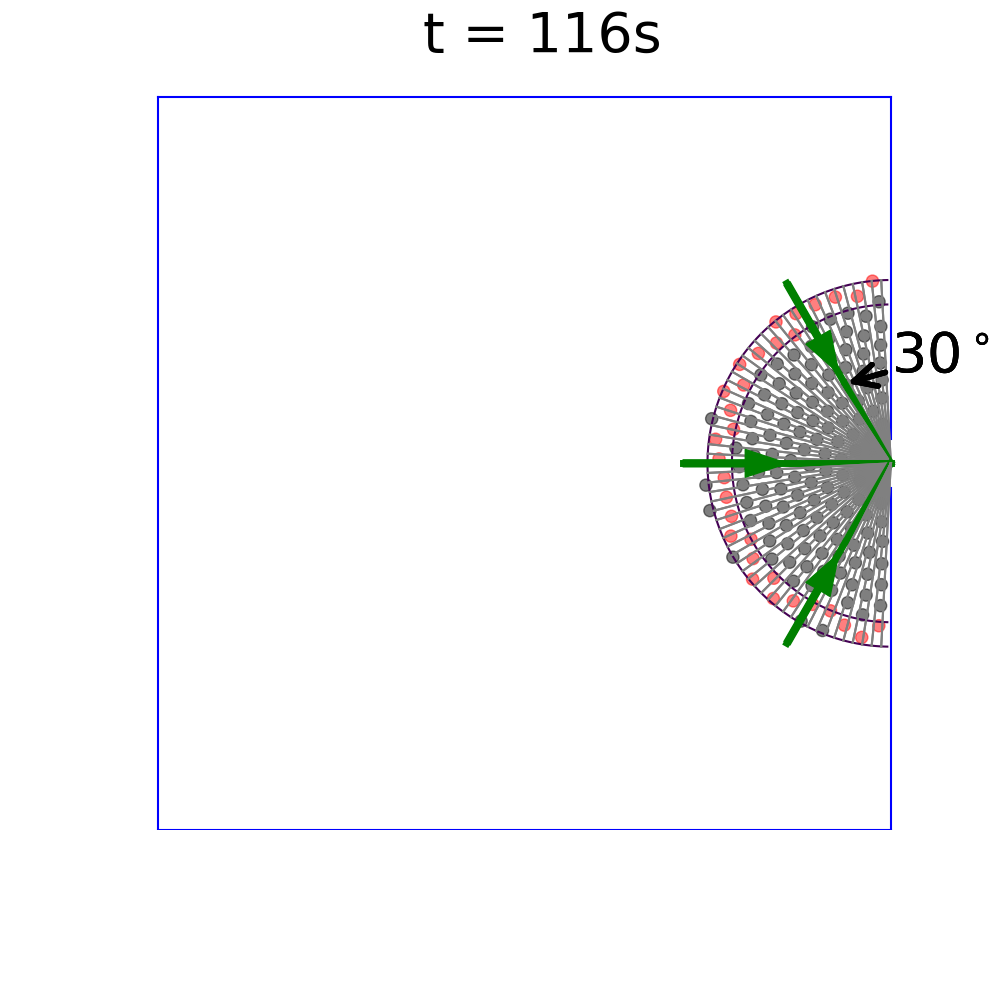}\\
    \vspace{-5mm}
    \includegraphics[width=\linewidth]{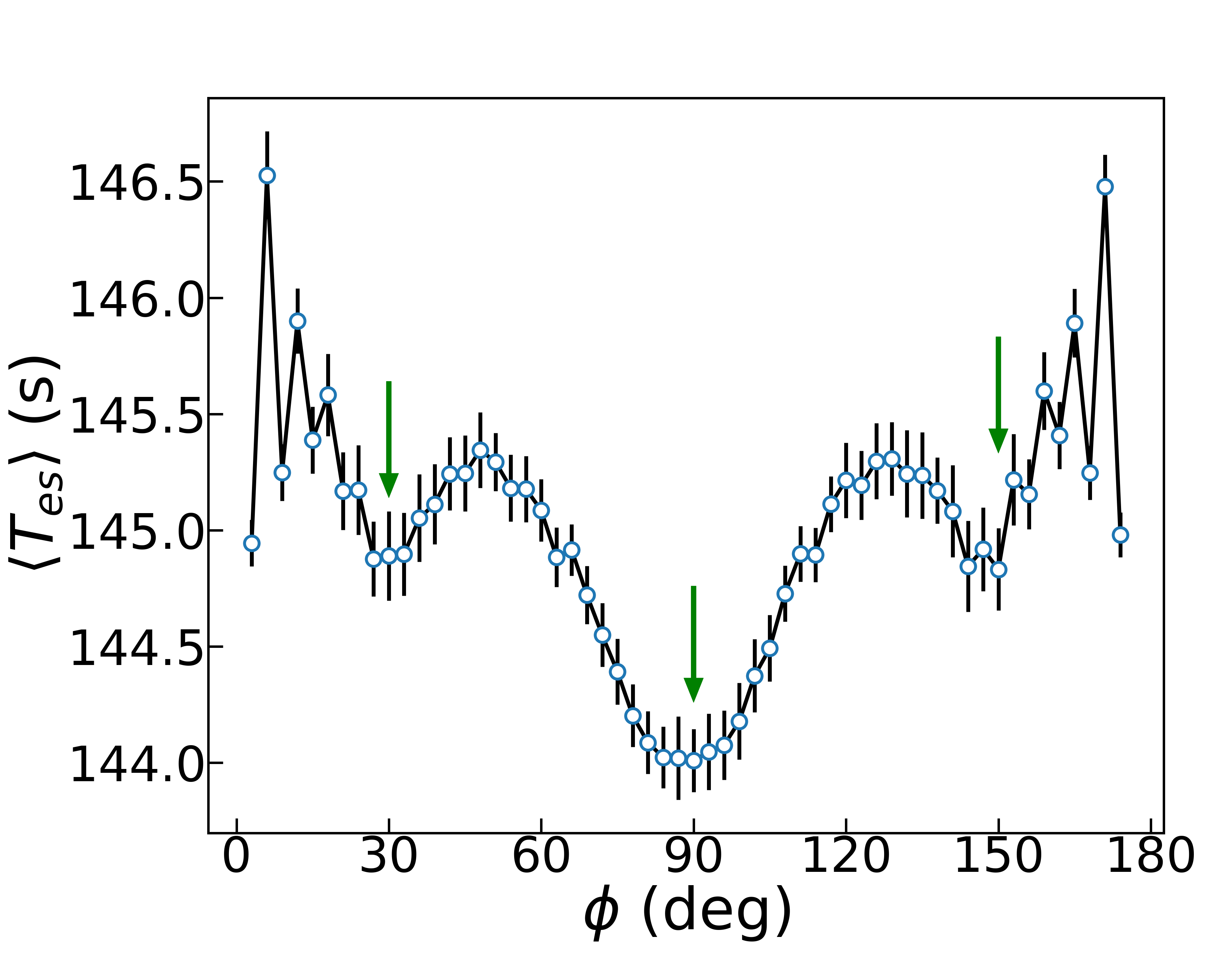}\\
  \end{minipage}
  \label{fig:3d}
}
\caption{Emergence of multiple fast tracks resulting from the
  spontaneously formed crystal structure in the escaping crowd. Typical
  snapshots of crowd configuration in the escape process are presented in the
  upper figures. The average escape time $\langle T_{es} \rangle$ of the
  pedestrians within the annulus between $R_1=13.5{\rm m}$ and $R_2=12.5{\rm m}$
  (indicated by red dots) is shown in the corresponding lower figures; the
  smaller annulus in $(d)$ is between $R_1=7.5{\rm m}$ and $R_2=6.5{\rm m}$;
  the averaging procedure is over 2000 independent simulations with random
  initial conditions and the error bars are obtained from 50 independent values
  of $\langle T_{es}(\phi) \rangle$. $\phi$ indicates the location of the
  pedestrian in the annulus. As the crowd is crystallised to a triangular
  lattice, valley structures emerge in the $\langle T_{es}(\phi) \rangle$ curve,
  which are identified as the fast tracks.}
\label{fig.3}
\end{figure*}

In this section we further explore the escape dynamics of the crowd by examining the motion of
individual pedestrians. In particular we are interested in this question: 
\textit{Where are the relatively safer spots in the collective
running of the pedestrians towards the exit?} To address this question, we focus
on a group of pedestrians within a narrow annulus around the exit, and track the
escape time of these pedestrians. The selected pedestrians are indicated by red dots,
as shown in Fig.~\ref{fig.3}. The distance between these pedestrians and the exit is approximately equal. 
In simulations, the annulus is created by drawing two adjacent circles centered at the door 
with the radii $R_1$ and $R_2$, respectively. The annulus is then equally divided into a number of zones. 
The spanning angle of each zone is specified to ensure that each zone is occupied by
pedestrians. We perform abundant independent simulations with random initial conditions.

By statistical analysis of the escape time of these pedestrians, we identify the
safe spots where pedestrians spend the least time to escape.  The lower panel in
Fig.~\ref{fig:3a} shows anisotropic feature in the escape of crowd. The average
escape time strongly depends on the angle $\phi$, which is defined in the upper
panel in Fig.~\ref{fig:3a}.  Remarkably, $\langle T_{es} \rangle$ reaches a deep
minimum at $\phi=90^{\circ}$. The average escape time of the pedestrians near
$\phi=90^{\circ}$ is only about a fourth of those near the wall.
Therefore, the pedestrians who are initially along the axis perpendicular to the
door spend significantly less time to escape than those near the wall. Note that
analysis of the empirical data of individual crowd escape events also shows the
anisotropy feature in the distributions of escape time, density and velocity in
typical room evacuations~\cite{garcimartin2017pedestrian,zuriguel2020contact}.

With the continuous outflow of the pedestrians, the crowd spontaneously form a
compact circular configuration, as shown in Fig.~\ref{fig:3b}-~\ref{fig:3d}.
Note that no clogging occurs at the exit in our system due to the relatively low
desired speed ($v^p = 1.0\,\rm{m/s}$); clogging may occur at high speed ($v^p >
1.5\,\rm{m/s}$)~\cite{Helbing2000VICSEK}.  The local minima developed on the
$\langle T_{es} \rangle$-$\phi$ curves in Fig.~\ref{fig:3b} -~\ref{fig:3d}
correspond to the relatively safe spots.  Remarkably, with the outflow of the
pedestrians, the trajectories of these safe spots constitute three straight
lines along the specific angles of $\phi=30^{\circ}$, $90^{\circ}$ and
$150^{\circ}$, as indicated by the green lines in the upper panels in
Fig.~\ref{fig.3}.  These lines are recognized as the fast tracks in the
collective escape of the crowd.

Here, it is natural to ask if the anisotropic feature of escape time originates
from the $\phi$-dependent trajectories of pedestrians. To check this point, we
examine the dependence of $\langle L \rangle /L_0$ on the angle $\phi$.
$\langle L \rangle$ is the mean actual length of the trajectory over independent
simulation runs, and $L_0$ is the length of the straight line connecting the
initial and final positions of a pedestrian at a specific angle $\phi$. The
actual trajectories of the pedestrians at $\phi=15^{\circ}$ and
$\phi=90^{\circ}$ are indicated by the green and red curves in Fig.~\ref{fig:zigzag}. 
The plot of $\langle L \rangle /L_0$ versus the angle $\phi$
is presented in Fig.~\ref{fig:M_shape}.  From the $\langle L \rangle /L_0$
curves, which correspond to the four cases in Fig.~\ref{fig.3}, we see that the local
minima do not match the angles of the fast tracks. As such, the emergence of the optimal 
angles could not be understood by the trajectory-based analysis. We are therefore led to 
examine the system from a new perspective.

\subsection{The perspective of crystallisation and topological defects}

The special angles of the fast tracks revealed in the preceding section
provide an important clue. The appearance of these discrete fast tracks may
be related to a global order developed in the compact packing of pedestrians.

\begin{figure*}[th]
\hspace{-6mm}
\subfigure[]{
  \begin{minipage}{0.33\textwidth}
    \centering
    \includegraphics[width=\linewidth]{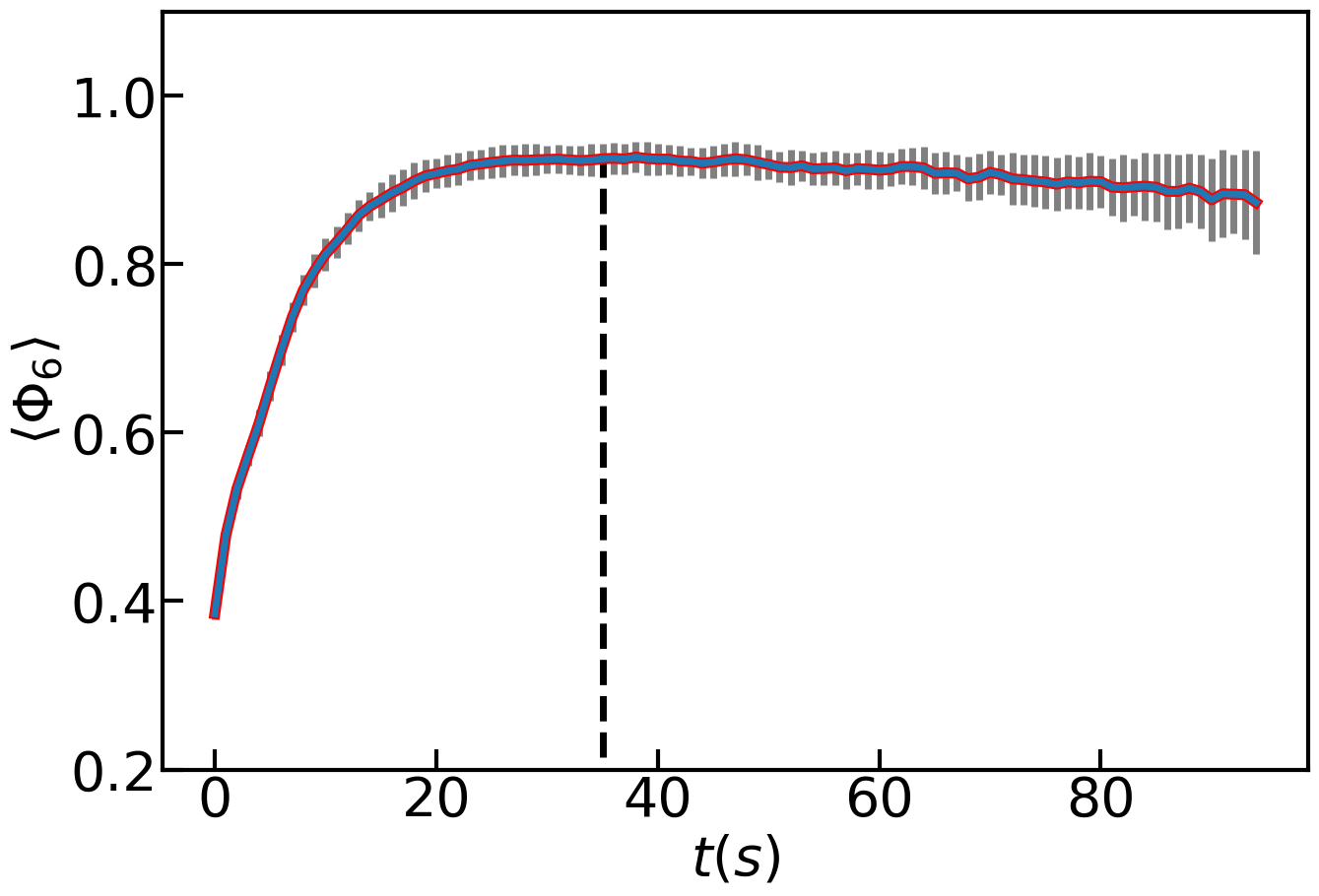}\\
    \vspace{-1mm}
  \end{minipage}
  \label{fig:4a}
}
\subfigure[]{
  \begin{minipage}{0.2\textwidth}
    \centering
    \includegraphics[width=\linewidth]{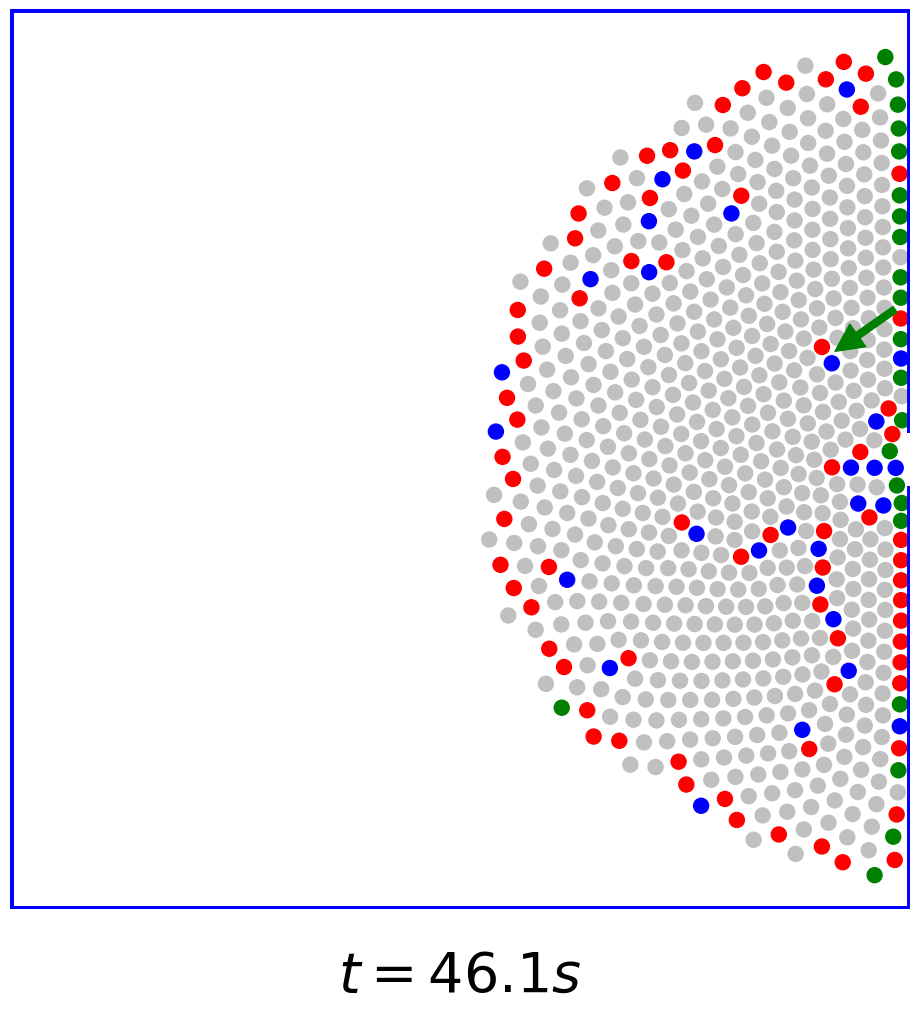}\\
  \end{minipage}
  \label{fig:4b}
}
\subfigure[]{
  \begin{minipage}{0.2\textwidth}
    \centering{}
    \includegraphics[width=\linewidth]{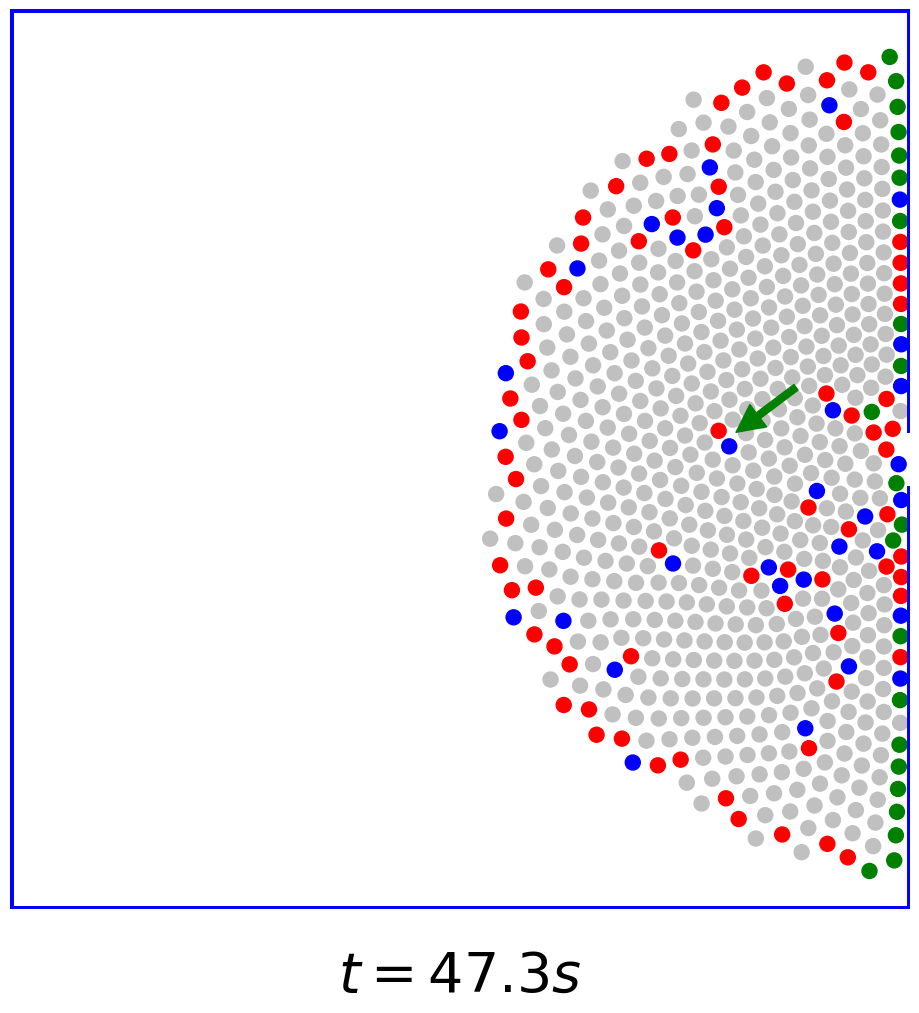}\\
  \end{minipage}
  \label{fig:4c}
}
\subfigure[]{
  \begin{minipage}{0.2\textwidth}
    \centering
    \includegraphics[width=\linewidth]{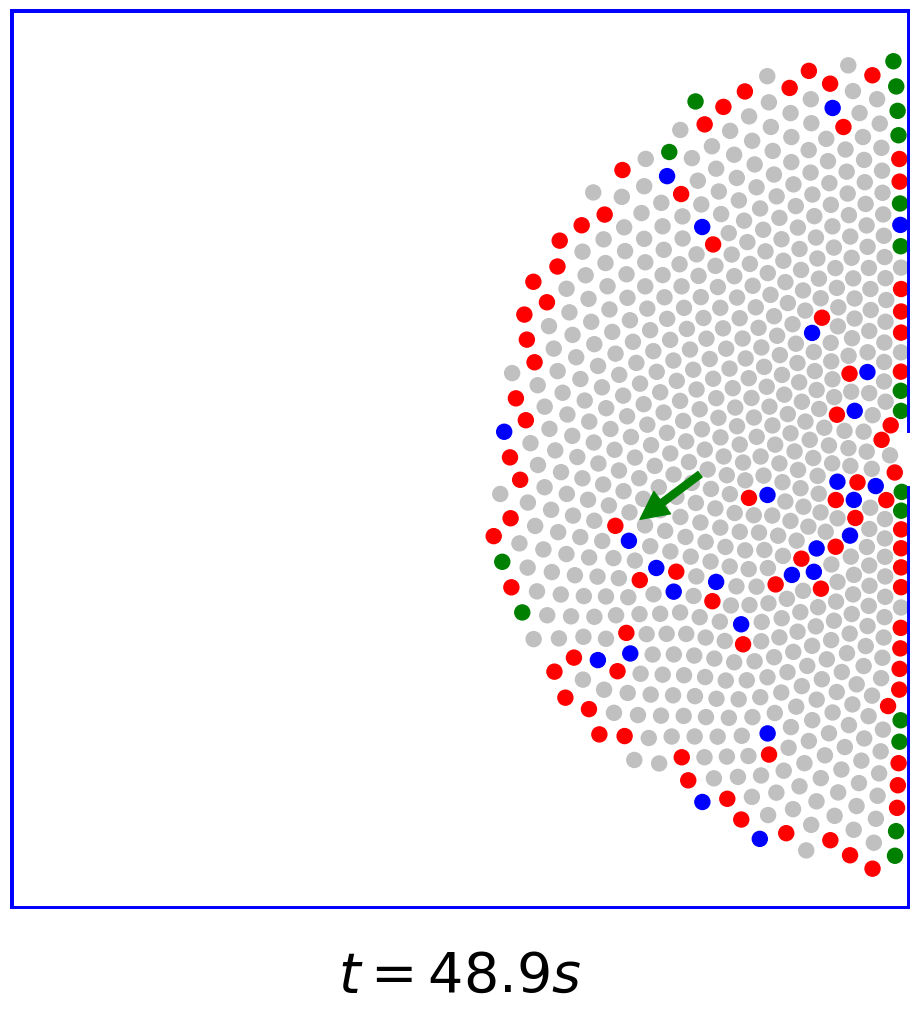}\\
  \end{minipage}{}
  \label{fig:4d}
}
{\caption{Crystallisation process in the collective escape. (a) Temporal
  variation of the six-fold bond-orientational order parameter $\langle \Phi_6
  \rangle$. The error bars are obtained from 100 independent simulations. The
  vertical dashed line indicates the time when the three fast tracks appear.
  (b)-(d) Glide motion of the disclocation as indicated by the green arrow in
  consecutive instantaneous configurations.  The red and blue dots represent
  five- and seven-fold disclinations, respectively. A dislocation consists of a
  pair of five- and seven-fold disclinations. }}
\label{fig.4}
\end{figure*}

\begin{figure}[th]
  \centering
  \hspace{4mm}
  \subfigure[]{\includegraphics[width=.34\textwidth]{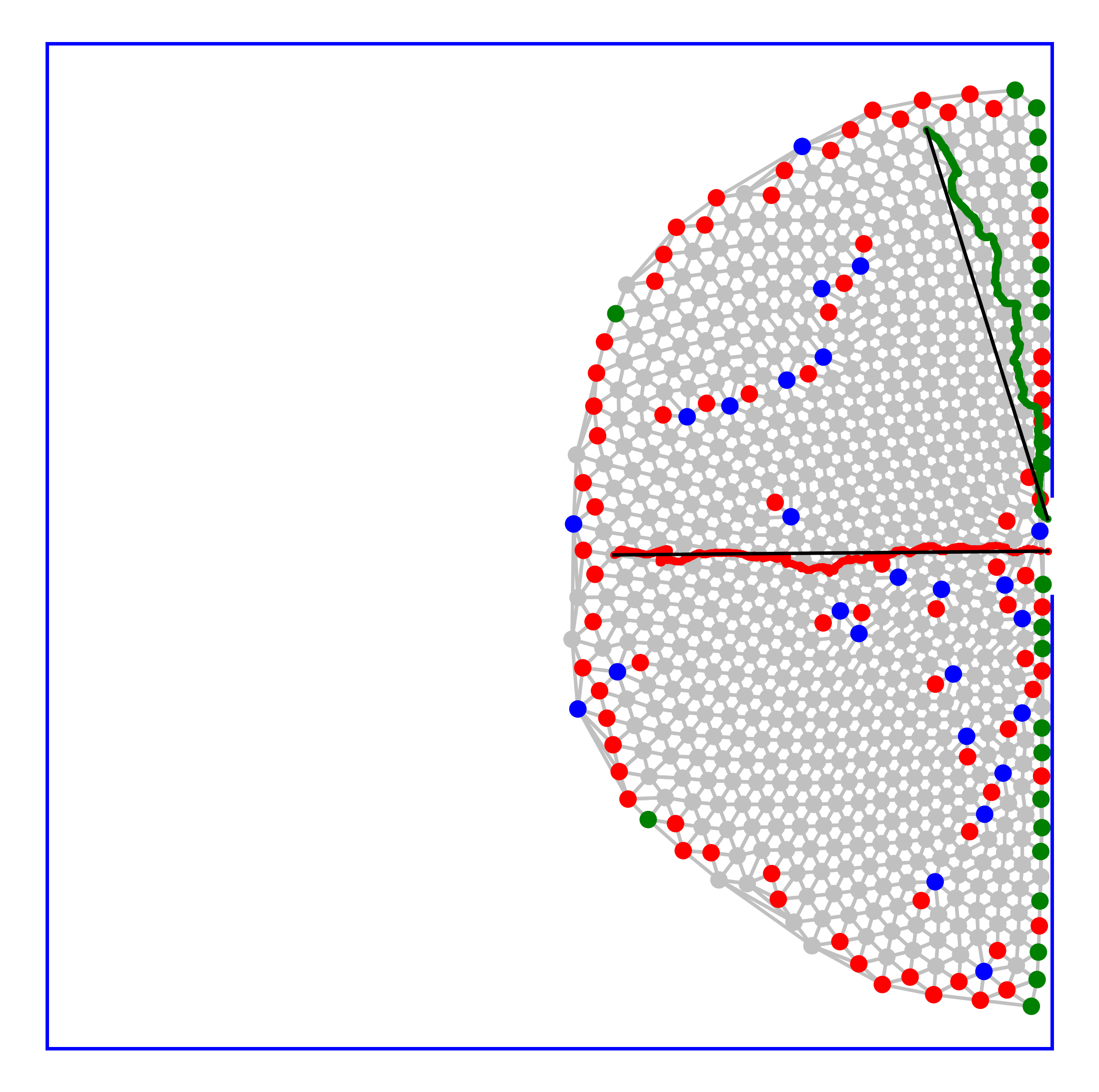}\label{fig:zigzag}}
  \subfigure[]{\includegraphics[width=.39\textwidth]{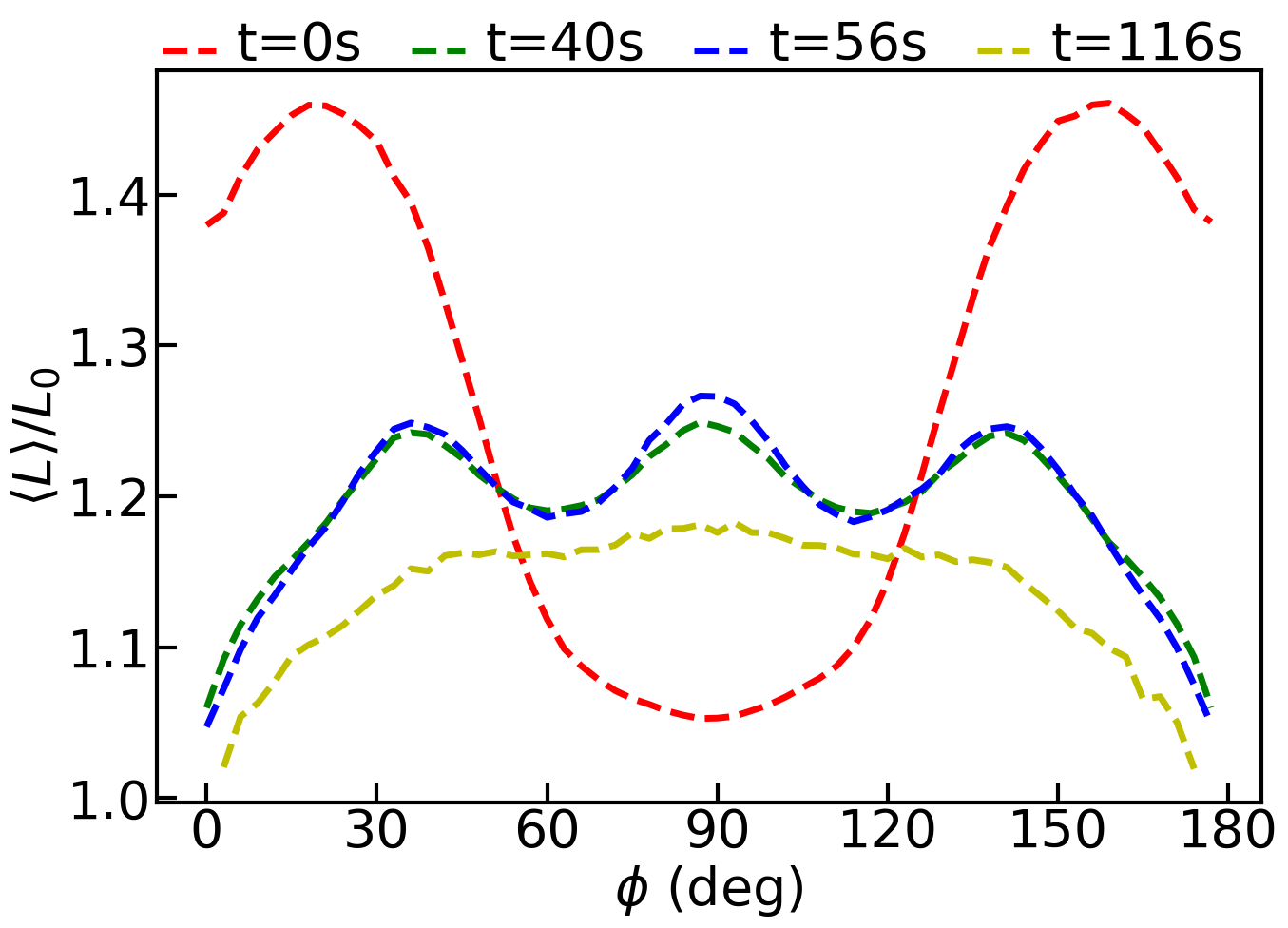}\label{fig:M_shape}}
  \caption{Statistical analysis of single pedestrian trajectories. 
    (a) Trajectories of typical pedestrians (see the red and green curves) in the crystallised
    crowd. Crystallographic defects are indicated by coloured dots. See the main text
    for more information. (b) Plot of $\langle L \rangle /L_0$ versus the angle $\phi$.
    $\langle L \rangle$ is the mean actual length of the trajectory over 2000 independent
    simulations with random initial conditions, and $L_0$ is the length of the straight line connecting the
    initial and final positions of a pedestrian at a specific angle $\phi$. } 
\label{fig.5}
\end{figure}

To reveal the positional order of the pedestrians, we treat each
pedestrian as a point and perform Delaunay triangulation of instantaneous crowd
configuration by establishing bonds between each pedestrian and the nearest
neighbours~\cite{chaikin2000principles,nelson2002defects}. This technique has
been widely used to quantitatively analyze crystalline
structures~\cite{chaikin2000principles, nelson2002defects}. An example of
Delaunay triangulation is shown in Fig.~\ref{fig:zigzag}. We see the
regular arrangement of pedestrians in the form of a 2D triangular lattice; each
vertex represents a pedestrian. Such a triangular lattice is disturbed by
defects as indicated by colored dots.  Specifically, the red and blue dots
represent pedestrians for whom the number of nearest neighbours (i.e., the
coordination number) deviates from six.  Note that in a perfect triangular
lattice, the coordination number for each vertex is always six. These coloured
dots therefore indicate the local disruption of the crystalline order.
They are known as topological defects, since they cannot be eliminated
by continuous deformation of the
medium~\cite{chaikin2000principles,nelson2002defects}. A vertex with
coordination number $5$ and $7$ are named five- and seven-fold disclinations,
respectively. These
disclinations are elementary topological defects in crystal lattice.

Interestingly, like electric charges, the disclinations carry topological charge
that reflects their intrinsic{} property. Specifically, the topological charge of
an $n$-fold disclination is $q=(6-n)\pi/3$. Charge $q$ is positive if $n < 6$ and
negative if $n > 6$.  According to elasticity theory, topological charges of the
same sign repel and unlike signs attract, which is analogous to electric
charges~\cite{chaikin2000principles,yao2014polydispersity}. The five-fold (red
dots) and seven-fold (blue dots) disclinations tend to form pairs known as
dislocations, as shown in Figs.~\ref{fig:4b}-\ref{fig:4d}.

In connection with our system, the complicated process of achieving
crystalline order in the compact crowd could be clarified by the concept of
topological defects. With the outflow of pedestrians, the microscopic
crystalline structure underlying the crowd is under persistent transformation.
Simulations show that this dynamical process exactly corresponds to the
characteristic annihilation and glide motion of topological defects in the
crystal. Annihilation of positive and negative topological charges reduces the
number of defects and leads to ordered arrangement of pedestrians. Furthermore,
dislocations are abundantly generated near the exit with the outflow of
pedestrians, and they tend to glide swiftly towards the boundary of the
crowd. These two
processes are crucial for achieving the crystallised state out of the initially
highly disordered configuration~\cite{yao2014polydispersity, yao2016dressed}.
In Figs.~\ref{fig:4b} -~\ref{fig:4d}, we show that the dislocation as
indicated by the green arrow swiftly glides along the direction that is
perpendicular to the line connecting the five- and seven-fold
disclinations~\cite{chaikin2000principles}. In this typical scenario of glide
motion, it takes about 2.7 seconds for the dislocation to move from the position
at Fig.~\ref{fig:4b} to that at Fig.~\ref{fig:4d}. In contrast, it takes about
8.4 seconds for a pedestrian of typical speed of $1 \rm{m/s}$ to cover the same
distance.

The crystallisation process could be characterized by the six-fold bond-orientational
order parameter~\cite{nelson2002defects}
\begin{eqnarray}\label{eq:6}
  \Phi_6 = \frac{1}{N} \sum_{i=1}^{N} | \frac{1}{n_b} \sum_{j=1}^{n_b}
  \exp(i6\theta_{ij}) |,
\end{eqnarray} 
where $n_b$ is the coordination number of the particle $i$, $\theta_{ij}$ is the
angle between the line connecting the particles $i$ and $j$ and some chosen
reference line, and $N$ is the total number of particles. Note that, to characterize
the interior crystalline order, a few layers of particles near the boundary and
the exit are excluded in the calculation for the order parameter.
The temporal variation of
$\langle \Phi_6 \rangle$ is presented in Fig.~\ref{fig:4a}; the error bars are obtained
by statistical analysis of 100 independent simulations. The vertical dashed
line indicates the time when the three fast tracks appear. This
observation indicates the strong connection of the emergence of the three fast
tracks and the full development of the crystalline order.

In the crystallised configuration of pedestrians, the principal axes of crystal
are invariant regardless of the microscopic motions of topological defects. In
fact, the transformation of the global crystalline structure which could modify the
orientations of the principal crystallographic axes, requires the appearance of
isolated disclinations; this is a highly energetically costly
process~\cite{yao2014polydispersity}. In simulations, we observe that the
orientations of the fast tracks are always along the three specific directions
in the entire evacuation process.  The central fast track that is
perpendicular to the wall emerges even before the crystalline order is
developed, as shown in Fig.~\ref{fig:3a}. And in general it does not correspond to the
principal axes of the crystal. The other two fast tracks make an angle of
$\pi/3$ with respect to the central one; this specific angle is a signature of
the triangular lattice. The stability of the global crystalline order well
explains the invariance of the orientations of the fast tracks. To conclude, the
formation of the fast tracks in the collective escape of human crowds has strong
connection with the global crystalline order developed in the configuration of
pedestrians.

\subsection{Effects of uncertainty in human behaviours}

As mentioned earlier, the motion of pedestrians is unambiguously determined by 
the physical and socio-psychological forces. However, the behaviours of pedestrians 
in a crowded environment could exhibit some degree of uncertainty as affected by
fluctuating psychological state and rapidly varying local environment. We model
the uncertainties in human behaviour by incorporating a random force into the
model. Here, we emphasize that adding a noise term to the original
deterministic SFM also provides an opportunity to test the robustness of the
model. The generalized social force model is
\begin{equation}\label{eq:7} 
m_i\frac{\rm{d} \boldsymbol{v}_i}{\rm{d} t} = m_i\frac{v_i^p(t)\boldsymbol{e}_i^p(t) -
\boldsymbol{v}_i(t)}{\tau_i} + \sum\limits_{j(\neq i)}\boldsymbol{f}_{ij} +
\sum\limits_{W}\boldsymbol{f}_{iW} + \boldsymbol{f}_{i}^N, 
\end{equation} 
in which the noise term $\boldsymbol{f}_{i}^N$ is given by 
\begin{eqnarray} \label{eq:8} 
\boldsymbol{f}_{i}^N = c\frac{mv^p}{\tau}\epsilon_i.
\end{eqnarray} 
The white noise $\epsilon_i$ follows a two-dimensional Gaussian distribution
${\rm N}(\boldsymbol{0}, {\rm I}_2)$. This approach takes into account the
flexible usage of space by pedestrians, which is essential to reproduce the
empirical observations in a natural and robust way ~\cite{helbing2005self}.  The
prefactor $c$ in eqn~(\ref{eq:8}) reflects the relative strength of
the noise force in comparison with the personal desire force. Note that
eqn (\ref{eq:7}) represents a particular active Brownian dynamics~\cite{schweitzer2007brownian}. 
The active force originates from the tendency for each
pedestrian to move towards the exit, and the noise term is for modelling the
uncertainties in human behaviour. While the values of the parameters in the
model are specified according to Ref.12 by comparison with experiments, the
variation of these parameters implies the richness in the collective dynamics of
the active agents. The investigation of eqn (\ref{eq:7}) as a model of active
Brownian dynamics in the parameter space is beyond the scope of this work.

To systematically investigate the effect of the random behaviour of pedestrians,
we vary the strength of the noise by tuning the value of $c$ in
eqn.(\ref{eq:8}). Crystallisation phenomenon is observed uniformly as the value
of $c$ is increased from 0.1 up to 100. The plot of $\langle T_{es} \rangle$
versus $\phi$ under varying $c$ is presented in Fig.~\ref{fig:noise}. For
reference, the noise-free case of $c=0$ is also plotted. Figure~\ref{fig:noise}
shows that adding a modest noise force does not qualitatively alter the
behaviours of the system, suggesting the robustness of the model. When the value
of $c$ is as large as 100, the crowd exhibits global migration which obscures
the concept of fast tracks.  We also notice that the collision of the crowd with
the wall leads to a density wave propagating through the entire system. See
Supplemental Materials for the videos of the simulated crowd dynamics at $c=0$,
$c=10$, and $c=100$ using the parameters in Table~\ref{tab.1}.

\subsection{Dilemma of escape strategy}

\begin{figure}[th]
  \centering
  \subfigure[]{\includegraphics[width=.39\textwidth]{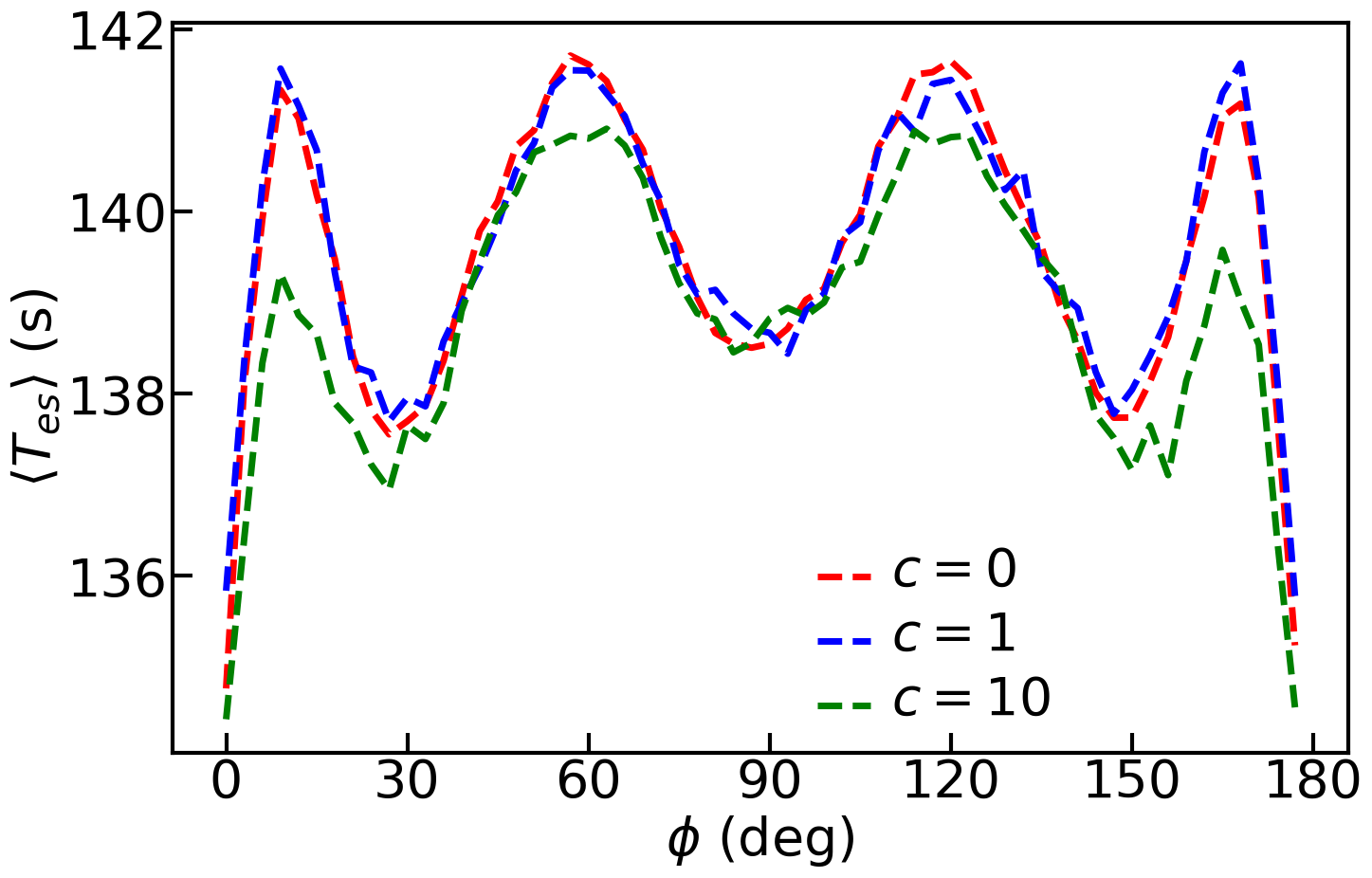}\label{fig:noise}}
  \subfigure[]{\includegraphics[width=.39\textwidth]{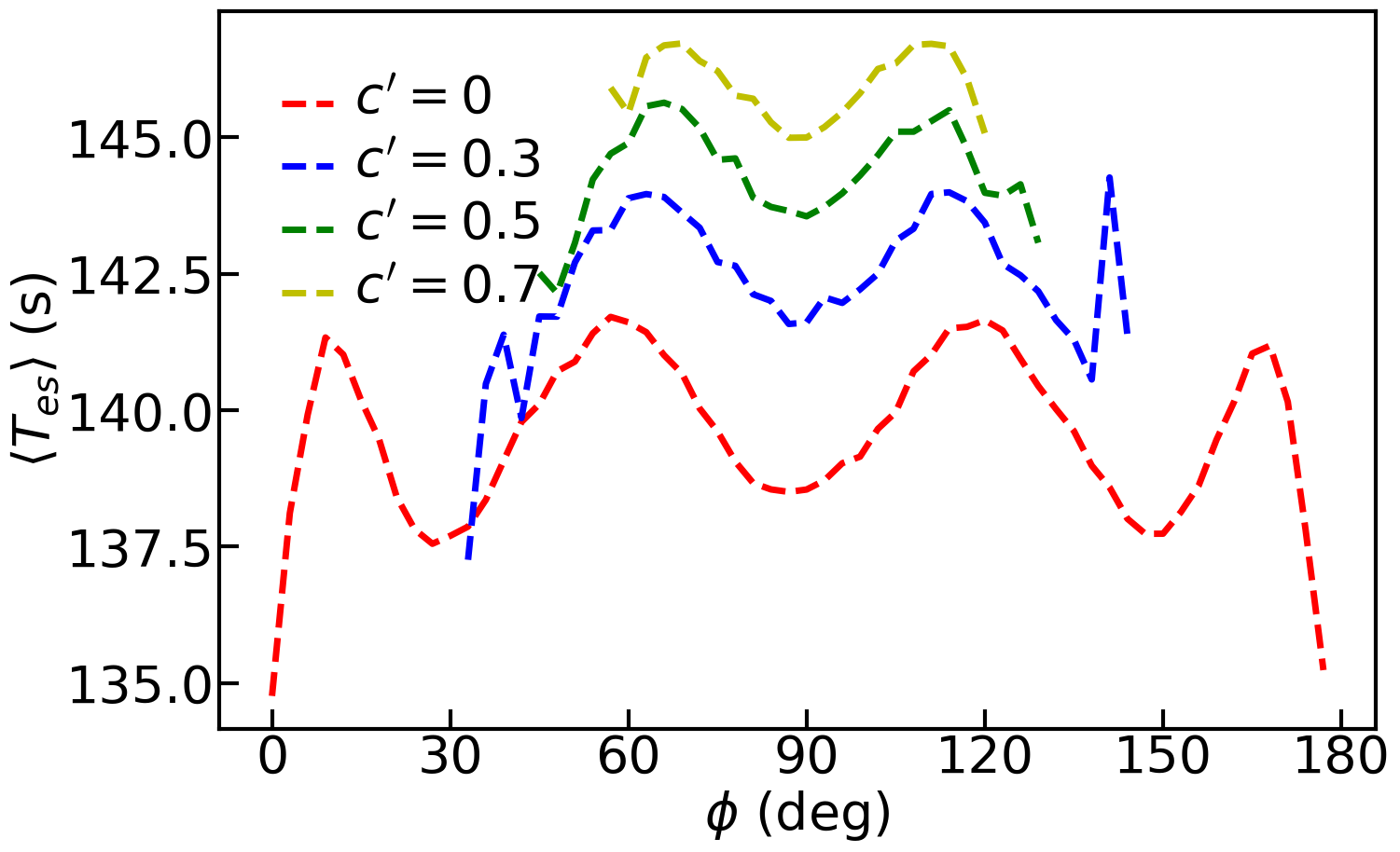}\label{fig:force}}
  \caption{\footnotesize Plot of the average escape time $\langle T_{es} \rangle$ of the
    pedestrians within the annulus over 2000 independent simulations 
    with random initial positions at varying conditions. The record of the escape time 
    starts at $t=56\,\rm{s}$ for comparison with Fig.~\ref{fig:3c}. (a) Increasing the
    uncertainty in decision-making, as modelled by the noise level in the model,
    does not affect the basic valley structures in the $\langle T_{es}(\phi) \rangle$ curve.
    (b) Including the tendency of pedestrians towards the central fast track boosts the 
    global risk. See the main text for more information.}
\label{fig.6}
\end{figure}

We proceed to discuss the scenario of collective escape if all the pedestrians know 
and adopt the strategy of running towards the \emph{central fast track}. To address this issue, 
we add a tendency force on each pedestrian towards the line of the central fast track, and 
update the crowd configuration by
\begin{equation}\label{eq:9} 
m_i\frac{\rm{d} \boldsymbol{v}_i}{\rm{d} t} = m_i\frac{v_i^p(t)\boldsymbol{e}_i^p(t) -
\boldsymbol{v}_i(t)}{\tau_i} + \sum\limits_{j(\neq i)}\boldsymbol{f}_{ij} +
\sum\limits_{W}\boldsymbol{f}_{iW} + \boldsymbol{f}_{i}^T. 
\end{equation} 
For simplicity, $\boldsymbol{f}_{i}^T$ is designed to be the gradient of a quadratic potential $U_i =
\frac{1}{2}bd_i^2$, where $d_i$ is the vertical distance from the pedestrian $i$ to the central horizontal line. 
The tendency force increases linearly with $d_i$. $b$ is the controlling parameter for the strength of the
tendency force. By writing $b$ in the form of $b =2c' mv^p/ H_0\tau$, the relative
strength of the tendency force is controlled by $c'$. The plot of $\langle T_{es} \rangle$ 
vs $\phi$ is presented in
Fig.~\ref{fig:force}. Increasing $c'$ leads to the elevation of the curves. It indicates
that when all the pedestrians adopt the escape strategy, the average evacuation
time increases. Nevertheless, the relatively safe spots are still located along the central
line; the relative advantage of the central line is reduced. \\

Finally, we emphasize that the results presented in this work are derived from
the theoretical SFM without considering relevant factors in reality, such as the
situation of injury and the intelligent adaption of pedestrians to local
environment. And we emphasize that the results derived in this work that
is based on the SFM are subject to experimental examination. Here,
we briefly discuss the social force model itself. This model has proven a
powerful tool to simulate complicated crowd dynamics by incorporating the
socio-psychological element into the force model.  However, the
socio-psychological forces in the model can be disputable. It is natural to
inquire to what extent the subtle psychology and action of pedestrians upon
various external stimuli can be characterised by these forces.  Our simulations
suggest that the emergent statistical law underlying the crowd dynamics is
unaffected by the modest random variation of individual behaviour.  This
observation provides good evidence on the reliability of the statistical laws
derived from the model.  But under larger fluctuations of individual behaviours,
such as non-local migration of pedestrians, one has to revise the forces. A
direction of interest is to design smart rules of motion that endow the
particles with more free will; the spirit of combining the physical and
socio-psychological elements may be retained in the new
model~\cite{imartinez2011multi}.

\section{Conclusion}

In summary, by conducting symmetric simulations with the social force model, we
have revealed the ordering driven multiple fast tracks in collective escape of
human crowds, and clarified the underlying mechanism by the exploiting the
striking analogy of crowd dynamics and crystallisation.  Notably, the emergent
topological defects and their characteristic dynamics play a crucial role for
shaping the global crystalline order and creating the fast tracks. 
It is important to emphasize that, it is not our intention to advocate any
specific model of crowd dynamics. 
Although our analysis in this work is based on the simulation results of SFM, we
conjecture, however, that the analysis methods proposed as well as the link
between crowd dynamics and crystallisation are not dependent on SFM. To this
end, it is of interest to explore such a link using other popular crowd dynamics
models~\cite{fiorini1998motion,ulicny2002towards} and further validate our analysis. Moreover, we appreciate that the
proposed mechanism interpretation needs to be further validated  by experimental
studies, and we believe that such mechanism may provide useful information
towards more efficient crowd evacuation management.

\section{ACKNOWLEDGMENTS}

This work was supported by NSFC Grants No. 16Z103010253.


\begin{thebibliography}{42}
\expandafter\ifx\csname natexlab\endcsname\relax\def\natexlab#1{#1}\fi
\expandafter\ifx\csname bibnamefont\endcsname\relax
  \def\bibnamefont#1{#1}\fi
\expandafter\ifx\csname bibfnamefont\endcsname\relax
  \def\bibfnamefont#1{#1}\fi
\expandafter\ifx\csname citenamefont\endcsname\relax
  \def\citenamefont#1{#1}\fi
\expandafter\ifx\csname url\endcsname\relax
  \def\url#1{\texttt{#1}}\fi
\expandafter\ifx\csname urlprefix\endcsname\relax\def\urlprefix{URL }\fi
\providecommand{\bibinfo}[2]{#2}
\providecommand{\eprint}[2][]{\url{#2}}

\bibitem[{\citenamefont{Helbing and Molnar}(1995)}]{Helbing1995Social}
\bibinfo{author}{\bibfnamefont{D.}~\bibnamefont{Helbing}} \bibnamefont{and}
  \bibinfo{author}{\bibfnamefont{P.}~\bibnamefont{Molnar}},
  \bibinfo{journal}{Phys. Rev. E} \textbf{\bibinfo{volume}{51}},
  \bibinfo{pages}{4282} (\bibinfo{year}{1995}).

\bibitem[{\citenamefont{Helbing et~al.}(2005)\citenamefont{Helbing, Buzna,
  Johansson, and Werner}}]{helbing2005self}
\bibinfo{author}{\bibfnamefont{D.}~\bibnamefont{Helbing}},
  \bibinfo{author}{\bibfnamefont{L.}~\bibnamefont{Buzna}},
  \bibinfo{author}{\bibfnamefont{A.}~\bibnamefont{Johansson}},
  \bibnamefont{and} \bibinfo{author}{\bibfnamefont{T.}~\bibnamefont{Werner}},
  \bibinfo{journal}{Transp. Sci.} \textbf{\bibinfo{volume}{39}},
  \bibinfo{pages}{1} (\bibinfo{year}{2005}).

\bibitem[{\citenamefont{Helbing et~al.}(2007)\citenamefont{Helbing, Johansson,
  and Al-Abideen}}]{helbing2007dynamics}
\bibinfo{author}{\bibfnamefont{D.}~\bibnamefont{Helbing}},
  \bibinfo{author}{\bibfnamefont{A.}~\bibnamefont{Johansson}},
  \bibnamefont{and} \bibinfo{author}{\bibfnamefont{H.~Z.}
  \bibnamefont{Al-Abideen}}, \bibinfo{journal}{Phys. Rev. E}
  \textbf{\bibinfo{volume}{75}}, \bibinfo{pages}{046109}
  (\bibinfo{year}{2007}).

\bibitem[{\citenamefont{Gravish et~al.}(2015)\citenamefont{Gravish, Gold,
  Zangwill, Goodisman, and Goldman}}]{gravish2015glass}
\bibinfo{author}{\bibfnamefont{N.}~\bibnamefont{Gravish}},
  \bibinfo{author}{\bibfnamefont{G.}~\bibnamefont{Gold}},
  \bibinfo{author}{\bibfnamefont{A.}~\bibnamefont{Zangwill}},
  \bibinfo{author}{\bibfnamefont{M.~A.} \bibnamefont{Goodisman}},
  \bibnamefont{and} \bibinfo{author}{\bibfnamefont{D.~I.}
  \bibnamefont{Goldman}}, \bibinfo{journal}{Soft matter}
  \textbf{\bibinfo{volume}{11}}, \bibinfo{pages}{6552} (\bibinfo{year}{2015}).

\bibitem[{\citenamefont{Tao and Dong}(2017)}]{tao2017floor}
\bibinfo{author}{\bibfnamefont{Y.~Z.} \bibnamefont{Tao}} \bibnamefont{and}
  \bibinfo{author}{\bibfnamefont{L.~Y.} \bibnamefont{Dong}},
  \bibinfo{journal}{Europhys. Lett.} \textbf{\bibinfo{volume}{119}},
  \bibinfo{pages}{10003} (\bibinfo{year}{2017}).

\bibitem[{\citenamefont{Rogsch et~al.}(2010)\citenamefont{Rogsch,
  Schreckenberg, Tribble, Klingsch, and Kretz}}]{rogsch2010panic}
\bibinfo{author}{\bibfnamefont{C.}~\bibnamefont{Rogsch}},
  \bibinfo{author}{\bibfnamefont{M.}~\bibnamefont{Schreckenberg}},
  \bibinfo{author}{\bibfnamefont{E.}~\bibnamefont{Tribble}},
  \bibinfo{author}{\bibfnamefont{W.}~\bibnamefont{Klingsch}}, \bibnamefont{and}
  \bibinfo{author}{\bibfnamefont{T.}~\bibnamefont{Kretz}}, in
  \emph{\bibinfo{booktitle}{Pedestrian and evacuation dynamics 2008}}
  (\bibinfo{publisher}{Springer Berlin Heidelberg}, \bibinfo{year}{2010}), pp.
  \bibinfo{pages}{743--755}.

\bibitem[{\citenamefont{Adrian et~al.}(2019)\citenamefont{Adrian, Amos,
  Baratchi, Beermann, Bode, Boltes, Corbetta, Dezecache, Drury, Fu
  et~al.}}]{adrian2019glossary}
\bibinfo{author}{\bibfnamefont{J.}~\bibnamefont{Adrian}},
  \bibinfo{author}{\bibfnamefont{M.}~\bibnamefont{Amos}},
  \bibinfo{author}{\bibfnamefont{M.}~\bibnamefont{Baratchi}},
  \bibinfo{author}{\bibfnamefont{M.}~\bibnamefont{Beermann}},
  \bibinfo{author}{\bibfnamefont{N.}~\bibnamefont{Bode}},
  \bibinfo{author}{\bibfnamefont{M.}~\bibnamefont{Boltes}},
  \bibinfo{author}{\bibfnamefont{A.}~\bibnamefont{Corbetta}},
  \bibinfo{author}{\bibfnamefont{G.}~\bibnamefont{Dezecache}},
  \bibinfo{author}{\bibfnamefont{J.}~\bibnamefont{Drury}},
  \bibinfo{author}{\bibfnamefont{Z.}~\bibnamefont{Fu}}, \bibnamefont{et~al.},
  \bibinfo{journal}{Collective Dynamics} \textbf{\bibinfo{volume}{4}},
  \bibinfo{pages}{1} (\bibinfo{year}{2019}).

\bibitem[{\citenamefont{Gibelli and Bellomo}(2019)}]{gibelli2019crowd}
\bibinfo{author}{\bibfnamefont{L.}~\bibnamefont{Gibelli}} \bibnamefont{and}
  \bibinfo{author}{\bibfnamefont{N.}~\bibnamefont{Bellomo}},
  \emph{\bibinfo{title}{Crowd Dynamics, Volume 1: Theory, Models, and Safety
  Problems}} (\bibinfo{publisher}{Springer}, \bibinfo{year}{2019}).

\bibitem[{\citenamefont{Rahouti et~al.}(2018)\citenamefont{Rahouti, Lovreglio,
  Jackson, and Datoussaid}}]{rahouti2018evacuation}
\bibinfo{author}{\bibfnamefont{A.}~\bibnamefont{Rahouti}},
  \bibinfo{author}{\bibfnamefont{R.}~\bibnamefont{Lovreglio}},
  \bibinfo{author}{\bibfnamefont{P.}~\bibnamefont{Jackson}}, \bibnamefont{and}
  \bibinfo{author}{\bibfnamefont{S.}~\bibnamefont{Datoussaid}}, in
  \emph{\bibinfo{booktitle}{Proceedings of the 9th International Conference on
  Pedestrian and Evacuation Dynamics (PED2018)}} (\bibinfo{year}{2018}).

\bibitem[{\citenamefont{Feliciani and
  Nishinari}(2018)}]{feliciani2018investigation}
\bibinfo{author}{\bibfnamefont{C.}~\bibnamefont{Feliciani}} \bibnamefont{and}
  \bibinfo{author}{\bibfnamefont{K.}~\bibnamefont{Nishinari}}, in
  \emph{\bibinfo{booktitle}{Proceedings of the 9th International Conference on
  Pedestrian and Evacuation Dynamics (PED2018)}} (\bibinfo{year}{2018}).

\bibitem[{\citenamefont{Smith and Dickie}(1993)}]{smith1993engineering}
\bibinfo{author}{\bibfnamefont{R.~A.} \bibnamefont{Smith}} \bibnamefont{and}
  \bibinfo{author}{\bibfnamefont{J.~F.} \bibnamefont{Dickie}},
  \emph{\bibinfo{title}{Engineering for crowd safety: proceedings of the
  International Conference on Engineering for Crowd Safety}}
  (\bibinfo{publisher}{Elsevier Science Ltd}, \bibinfo{year}{1993}).

\bibitem[{\citenamefont{Helbing et~al.}(2000)\citenamefont{Helbing, Farkas, and
  Vicsek}}]{Helbing2000VICSEK}
\bibinfo{author}{\bibfnamefont{D.}~\bibnamefont{Helbing}},
  \bibinfo{author}{\bibfnamefont{I.}~\bibnamefont{Farkas}}, \bibnamefont{and}
  \bibinfo{author}{\bibfnamefont{T.}~\bibnamefont{Vicsek}},
  \bibinfo{journal}{Nature} \textbf{\bibinfo{volume}{407}},
  \bibinfo{pages}{487} (\bibinfo{year}{2000}).

\bibitem[{\citenamefont{Batty et~al.}(2003)\citenamefont{Batty, DeSyllas, and
  Duxbury}}]{batty2003discrete}
\bibinfo{author}{\bibfnamefont{M.}~\bibnamefont{Batty}},
  \bibinfo{author}{\bibfnamefont{J.}~\bibnamefont{DeSyllas}}, \bibnamefont{and}
  \bibinfo{author}{\bibfnamefont{E.}~\bibnamefont{Duxbury}},
  \bibinfo{journal}{Int. J. Geogr. Inf. Sci.} \textbf{\bibinfo{volume}{17}},
  \bibinfo{pages}{673} (\bibinfo{year}{2003}).

\bibitem[{\citenamefont{Metivet et~al.}(2018)\citenamefont{Metivet, Pastorello,
  and Peyla}}]{metivet2018how}
\bibinfo{author}{\bibfnamefont{T.}~\bibnamefont{Metivet}},
  \bibinfo{author}{\bibfnamefont{L.}~\bibnamefont{Pastorello}},
  \bibnamefont{and} \bibinfo{author}{\bibfnamefont{P.}~\bibnamefont{Peyla}},
  \bibinfo{journal}{Europhys. Lett.} \textbf{\bibinfo{volume}{121}},
  \bibinfo{pages}{54003} (\bibinfo{year}{2018}).

\bibitem[{\citenamefont{Moussa{\"\i}d et~al.}(2010)\citenamefont{Moussa{\"\i}d,
  Perozo, Garnier, Helbing, and Theraulaz}}]{moussaid2010walking}
\bibinfo{author}{\bibfnamefont{M.}~\bibnamefont{Moussa{\"\i}d}},
  \bibinfo{author}{\bibfnamefont{N.}~\bibnamefont{Perozo}},
  \bibinfo{author}{\bibfnamefont{S.}~\bibnamefont{Garnier}},
  \bibinfo{author}{\bibfnamefont{D.}~\bibnamefont{Helbing}}, \bibnamefont{and}
  \bibinfo{author}{\bibfnamefont{G.}~\bibnamefont{Theraulaz}},
  \bibinfo{journal}{PloS one} \textbf{\bibinfo{volume}{5}},
  \bibinfo{pages}{e10047} (\bibinfo{year}{2010}).

\bibitem[{\citenamefont{James}(1953)}]{james1953distribution}
\bibinfo{author}{\bibfnamefont{J.}~\bibnamefont{James}}, \bibinfo{journal}{Am.
  Sociol. Rev.} \textbf{\bibinfo{volume}{18}}, \bibinfo{pages}{569}
  (\bibinfo{year}{1953}).

\bibitem[{\citenamefont{Sumpter}(2006)}]{sumpter2006principles}
\bibinfo{author}{\bibfnamefont{D.~J.} \bibnamefont{Sumpter}},
  \bibinfo{journal}{Philos. Trans. R. Soc. B-Biol. Sci.}
  \textbf{\bibinfo{volume}{361}}, \bibinfo{pages}{5} (\bibinfo{year}{2006}).

\bibitem[{\citenamefont{Castellano et~al.}(2009)\citenamefont{Castellano,
  Fortunato, and Loreto}}]{Castellano2009Statistical}
\bibinfo{author}{\bibfnamefont{C.}~\bibnamefont{Castellano}},
  \bibinfo{author}{\bibfnamefont{S.}~\bibnamefont{Fortunato}},
  \bibnamefont{and} \bibinfo{author}{\bibfnamefont{V.}~\bibnamefont{Loreto}},
  \bibinfo{journal}{Rev. Mod. Phys.} \textbf{\bibinfo{volume}{81}},
  \bibinfo{pages}{591} (\bibinfo{year}{2009}).

\bibitem[{\citenamefont{Schadschneider
  et~al.}(2009)\citenamefont{Schadschneider, Klingsch, Kl{\"u}pfel, Kretz,
  Rogsch, and Seyfried}}]{schadschneider2009evacuation}
\bibinfo{author}{\bibfnamefont{A.}~\bibnamefont{Schadschneider}},
  \bibinfo{author}{\bibfnamefont{W.}~\bibnamefont{Klingsch}},
  \bibinfo{author}{\bibfnamefont{H.}~\bibnamefont{Kl{\"u}pfel}},
  \bibinfo{author}{\bibfnamefont{T.}~\bibnamefont{Kretz}},
  \bibinfo{author}{\bibfnamefont{C.}~\bibnamefont{Rogsch}}, \bibnamefont{and}
  \bibinfo{author}{\bibfnamefont{A.}~\bibnamefont{Seyfried}},
  \emph{\bibinfo{title}{Evacuation Dynamics: Empirical Results, Modeling and
  Applications}} (\bibinfo{publisher}{Springer}, \bibinfo{address}{New York,
  NY}, \bibinfo{year}{2009}), pp. \bibinfo{pages}{3142--3176},
  \bibinfo{edition}{1st} ed.

\bibitem[{\citenamefont{Lakoba et~al.}(2005)\citenamefont{Lakoba, Kaup, and
  Finkelstein}}]{lakoba2005modifications}
\bibinfo{author}{\bibfnamefont{T.~I.} \bibnamefont{Lakoba}},
  \bibinfo{author}{\bibfnamefont{D.~J.} \bibnamefont{Kaup}}, \bibnamefont{and}
  \bibinfo{author}{\bibfnamefont{N.~M.} \bibnamefont{Finkelstein}},
  \bibinfo{journal}{Simulation} \textbf{\bibinfo{volume}{81}},
  \bibinfo{pages}{339} (\bibinfo{year}{2005}).

\bibitem[{\citenamefont{Zuriguel et~al.}(2020)\citenamefont{Zuriguel,
  Echeverria, Maza, Hidalgo, Mart{\'\i}n-G{\'o}mez, and
  Garcimart{\'\i}n}}]{zuriguel2020contact}
\bibinfo{author}{\bibfnamefont{I.}~\bibnamefont{Zuriguel}},
  \bibinfo{author}{\bibfnamefont{I.}~\bibnamefont{Echeverria}},
  \bibinfo{author}{\bibfnamefont{D.}~\bibnamefont{Maza}},
  \bibinfo{author}{\bibfnamefont{R.~C.} \bibnamefont{Hidalgo}},
  \bibinfo{author}{\bibfnamefont{C.}~\bibnamefont{Mart{\'\i}n-G{\'o}mez}},
  \bibnamefont{and}
  \bibinfo{author}{\bibfnamefont{A.}~\bibnamefont{Garcimart{\'\i}n}},
  \bibinfo{journal}{Saf. Sci.} \textbf{\bibinfo{volume}{121}},
  \bibinfo{pages}{394} (\bibinfo{year}{2020}).

\bibitem[{\citenamefont{Chraibi et~al.}(2015)\citenamefont{Chraibi, Ezaki,
  Tordeux, Nishinari, Schadschneider, and Seyfried}}]{chraibi2015jamming}
\bibinfo{author}{\bibfnamefont{M.}~\bibnamefont{Chraibi}},
  \bibinfo{author}{\bibfnamefont{T.}~\bibnamefont{Ezaki}},
  \bibinfo{author}{\bibfnamefont{A.}~\bibnamefont{Tordeux}},
  \bibinfo{author}{\bibfnamefont{K.}~\bibnamefont{Nishinari}},
  \bibinfo{author}{\bibfnamefont{A.}~\bibnamefont{Schadschneider}},
  \bibnamefont{and} \bibinfo{author}{\bibfnamefont{A.}~\bibnamefont{Seyfried}},
  \bibinfo{journal}{Phys. Rev. E} \textbf{\bibinfo{volume}{92}},
  \bibinfo{pages}{042809} (\bibinfo{year}{2015}).

\bibitem[{\citenamefont{Boltes et~al.}(2018)\citenamefont{Boltes, Zhang,
  Tordeux, Schadschneider, and Seyfried}}]{boltes2018empirical}
\bibinfo{author}{\bibfnamefont{M.}~\bibnamefont{Boltes}},
  \bibinfo{author}{\bibfnamefont{J.}~\bibnamefont{Zhang}},
  \bibinfo{author}{\bibfnamefont{A.}~\bibnamefont{Tordeux}},
  \bibinfo{author}{\bibfnamefont{A.}~\bibnamefont{Schadschneider}},
  \bibnamefont{and} \bibinfo{author}{\bibfnamefont{A.}~\bibnamefont{Seyfried}},
  \emph{\bibinfo{title}{Empirical Results of Pedestrian and Evacuation
  Dynamics}} (\bibinfo{publisher}{Springer}, \bibinfo{address}{Berlin,
  Heidelberg}, \bibinfo{year}{2018}), pp. \bibinfo{pages}{1--29},
  \bibinfo{edition}{2nd} ed.

\bibitem[{\citenamefont{Cordes et~al.}(2019)\citenamefont{Cordes,
  Schadschneider, and Tordeux}}]{cordes2019trouble}
\bibinfo{author}{\bibfnamefont{J.}~\bibnamefont{Cordes}},
  \bibinfo{author}{\bibfnamefont{A.}~\bibnamefont{Schadschneider}},
  \bibnamefont{and} \bibinfo{author}{\bibfnamefont{A.}~\bibnamefont{Tordeux}},
  \bibinfo{journal}{arXiv preprint arXiv:1911.07547}  (\bibinfo{year}{2019}).

\bibitem[{\citenamefont{Helbing and Johansson}(2009)}]{helbing2009pedestrian}
\bibinfo{author}{\bibfnamefont{D.}~\bibnamefont{Helbing}} \bibnamefont{and}
  \bibinfo{author}{\bibfnamefont{A.}~\bibnamefont{Johansson}},
  \emph{\bibinfo{title}{Pedestrian, Crowd and Evacuation Dynamics}}
  (\bibinfo{publisher}{Springer}, \bibinfo{address}{New York, NY},
  \bibinfo{year}{2009}), pp. \bibinfo{pages}{6476--6495},
  \bibinfo{edition}{1st} ed.

\bibitem[{\citenamefont{Henderson}(1971)}]{henderson1971statistics}
\bibinfo{author}{\bibfnamefont{L.}~\bibnamefont{Henderson}},
  \bibinfo{journal}{Nature} \textbf{\bibinfo{volume}{229}},
  \bibinfo{pages}{381} (\bibinfo{year}{1971}).

\bibitem[{\citenamefont{Henderson}(1974)}]{henderson1974fluid}
\bibinfo{author}{\bibfnamefont{L.~F.} \bibnamefont{Henderson}},
  \bibinfo{journal}{Transportation Research} \textbf{\bibinfo{volume}{8}},
  \bibinfo{pages}{509} (\bibinfo{year}{1974}).

\bibitem[{\citenamefont{Helbing}(1998)}]{helbing1998fluid}
\bibinfo{author}{\bibfnamefont{D.}~\bibnamefont{Helbing}},
  \bibinfo{journal}{Complex Systems} \textbf{\bibinfo{volume}{6}},
  \bibinfo{pages}{391} (\bibinfo{year}{1998}).

\bibitem[{\citenamefont{Bauer et~al.}(2007)\citenamefont{Bauer, Seer, and
  Br{\"a}ndle}}]{bauer2007macroscopic}
\bibinfo{author}{\bibfnamefont{D.}~\bibnamefont{Bauer}},
  \bibinfo{author}{\bibfnamefont{S.}~\bibnamefont{Seer}}, \bibnamefont{and}
  \bibinfo{author}{\bibfnamefont{N.}~\bibnamefont{Br{\"a}ndle}}, in
  \emph{\bibinfo{booktitle}{Proceedings of the 2007 summer computer simulation
  conference}} (\bibinfo{organization}{Society for Computer Simulation
  International}, \bibinfo{address}{San Diego, CA, USA}, \bibinfo{year}{2007}),
  pp. \bibinfo{pages}{1035--1042}.

\bibitem[{\citenamefont{Helbing et~al.}(2003)\citenamefont{Helbing, Isobe,
  Nagatani, and Takimoto}}]{helbing2003lattice}
\bibinfo{author}{\bibfnamefont{D.}~\bibnamefont{Helbing}},
  \bibinfo{author}{\bibfnamefont{M.}~\bibnamefont{Isobe}},
  \bibinfo{author}{\bibfnamefont{T.}~\bibnamefont{Nagatani}}, \bibnamefont{and}
  \bibinfo{author}{\bibfnamefont{K.}~\bibnamefont{Takimoto}},
  \bibinfo{journal}{Phys. Rev. E} \textbf{\bibinfo{volume}{67}},
  \bibinfo{pages}{067101} (\bibinfo{year}{2003}).

\bibitem[{\citenamefont{Czirók and Vicsek}(2012)}]{Czir2012Collective}
\bibinfo{author}{\bibfnamefont{A.}~\bibnamefont{Czirók}} \bibnamefont{and}
  \bibinfo{author}{\bibfnamefont{T.}~\bibnamefont{Vicsek}},
  \bibinfo{journal}{Physica A} \textbf{\bibinfo{volume}{281}},
  \bibinfo{pages}{17} (\bibinfo{year}{2012}).

\bibitem[{\citenamefont{Chraibi et~al.}(2018)\citenamefont{Chraibi, Tordeux,
  Schadschneider, and Seyfried}}]{chraibi2018modelling}
\bibinfo{author}{\bibfnamefont{M.}~\bibnamefont{Chraibi}},
  \bibinfo{author}{\bibfnamefont{A.}~\bibnamefont{Tordeux}},
  \bibinfo{author}{\bibfnamefont{A.}~\bibnamefont{Schadschneider}},
  \bibnamefont{and} \bibinfo{author}{\bibfnamefont{A.}~\bibnamefont{Seyfried}},
  \emph{\bibinfo{title}{Modelling of Pedestrian and Evacuation Dynamics}}
  (\bibinfo{publisher}{Springer}, \bibinfo{address}{Berlin, Heidelberg},
  \bibinfo{year}{2018}), pp. \bibinfo{pages}{1--22}, \bibinfo{edition}{2nd} ed.

\bibitem[{\citenamefont{Corbetta et~al.}(2018)\citenamefont{Corbetta, Meeusen,
  Lee, Benzi, and Toschi}}]{corbetta2018physics}
\bibinfo{author}{\bibfnamefont{A.}~\bibnamefont{Corbetta}},
  \bibinfo{author}{\bibfnamefont{J.~A.} \bibnamefont{Meeusen}},
  \bibinfo{author}{\bibfnamefont{C.-m.} \bibnamefont{Lee}},
  \bibinfo{author}{\bibfnamefont{R.}~\bibnamefont{Benzi}}, \bibnamefont{and}
  \bibinfo{author}{\bibfnamefont{F.}~\bibnamefont{Toschi}},
  \bibinfo{journal}{Phys. Rev. E} \textbf{\bibinfo{volume}{98}},
  \bibinfo{pages}{062310} (\bibinfo{year}{2018}).

\bibitem[{\citenamefont{Das et~al.}(2020)\citenamefont{Das, Schmidt, and
  Murrell}}]{das2020introduction}
\bibinfo{author}{\bibfnamefont{M.}~\bibnamefont{Das}},
  \bibinfo{author}{\bibfnamefont{C.~F.} \bibnamefont{Schmidt}},
  \bibnamefont{and} \bibinfo{author}{\bibfnamefont{M.}~\bibnamefont{Murrell}},
  \bibinfo{journal}{Soft Matter} \textbf{\bibinfo{volume}{16}},
  \bibinfo{pages}{7185} (\bibinfo{year}{2020}).

\bibitem[{\citenamefont{Helbing}(1991)}]{helbing1991mathematical}
\bibinfo{author}{\bibfnamefont{D.}~\bibnamefont{Helbing}},
  \bibinfo{journal}{Behavioral Sci.} \textbf{\bibinfo{volume}{36}},
  \bibinfo{pages}{298} (\bibinfo{year}{1991}).

\bibitem[{\citenamefont{Colombi and Scianna}(2017)}]{Colombi2017Modelling}
\bibinfo{author}{\bibfnamefont{A.}~\bibnamefont{Colombi}} \bibnamefont{and}
  \bibinfo{author}{\bibfnamefont{M.}~\bibnamefont{Scianna}},
  \bibinfo{journal}{R. Soc. Open Sci.} \textbf{\bibinfo{volume}{4}},
  \bibinfo{pages}{160561} (\bibinfo{year}{2017}).

\bibitem[{\citenamefont{Nelson}(2002)}]{nelson2002defects}
\bibinfo{author}{\bibfnamefont{D.~R.} \bibnamefont{Nelson}},
  \emph{\bibinfo{title}{Defects and geometry in condensed matter physics}}
  (\bibinfo{publisher}{Cambridge University Press},
  \bibinfo{address}{Cambridge, UK}, \bibinfo{year}{2002}).

\bibitem[{\citenamefont{Yao and Olvera de~la
  Cruz}(2014)}]{yao2014polydispersity}
\bibinfo{author}{\bibfnamefont{Z.}~\bibnamefont{Yao}} \bibnamefont{and}
  \bibinfo{author}{\bibfnamefont{M.}~\bibnamefont{Olvera de~la Cruz}},
  \bibinfo{journal}{Proc. Natl. Acad. Sci. USA} \textbf{\bibinfo{volume}{111}},
  \bibinfo{pages}{5094} (\bibinfo{year}{2014}).

\bibitem[{\citenamefont{Yao}(2016)}]{yao2016dressed}
\bibinfo{author}{\bibfnamefont{Z.}~\bibnamefont{Yao}}, \bibinfo{journal}{Soft
  Matter} \textbf{\bibinfo{volume}{12}}, \bibinfo{pages}{7020}
  (\bibinfo{year}{2016}).

\bibitem[{\citenamefont{Chaikin and Lubensky}(2000)}]{chaikin2000principles}
\bibinfo{author}{\bibfnamefont{P.}~\bibnamefont{Chaikin}} \bibnamefont{and}
  \bibinfo{author}{\bibfnamefont{T.}~\bibnamefont{Lubensky}},
  \emph{\bibinfo{title}{Principles of condensed matter physics}}
  (\bibinfo{publisher}{Cambridge University Press},
  \bibinfo{address}{Cambridge, UK}, \bibinfo{year}{2000}).

\bibitem[{\citenamefont{Matin et~al.}(2003)\citenamefont{Matin, Daivis, and
  Todd}}]{matin2003cell}
\bibinfo{author}{\bibfnamefont{M.}~\bibnamefont{Matin}},
  \bibinfo{author}{\bibfnamefont{P.}~\bibnamefont{Daivis}}, \bibnamefont{and}
  \bibinfo{author}{\bibfnamefont{B.}~\bibnamefont{Todd}},
  \bibinfo{journal}{Comput. Phys. Commun.} \textbf{\bibinfo{volume}{151}},
  \bibinfo{pages}{35} (\bibinfo{year}{2003}).

\bibitem[{\citenamefont{Dobson et~al.}(2016)\citenamefont{Dobson, Fox, and
  Saracino}}]{dobson2016cell}
\bibinfo{author}{\bibfnamefont{M.}~\bibnamefont{Dobson}},
  \bibinfo{author}{\bibfnamefont{I.}~\bibnamefont{Fox}}, \bibnamefont{and}
  \bibinfo{author}{\bibfnamefont{A.}~\bibnamefont{Saracino}},
  \bibinfo{journal}{J. Comput. Phys.} \textbf{\bibinfo{volume}{315}},
  \bibinfo{pages}{211} (\bibinfo{year}{2016}).

\bibitem[{\citenamefont{Garcimart{\'\i}n
  et~al.}(2017)\citenamefont{Garcimart{\'\i}n, Pastor, Mart{\'\i}n-G{\'o}mez,
  Parisi, and Zuriguel}}]{garcimartin2017pedestrian}
\bibinfo{author}{\bibfnamefont{A.}~\bibnamefont{Garcimart{\'\i}n}},
  \bibinfo{author}{\bibfnamefont{J.~M.} \bibnamefont{Pastor}},
  \bibinfo{author}{\bibfnamefont{C.}~\bibnamefont{Mart{\'\i}n-G{\'o}mez}},
  \bibinfo{author}{\bibfnamefont{D.}~\bibnamefont{Parisi}}, \bibnamefont{and}
  \bibinfo{author}{\bibfnamefont{I.}~\bibnamefont{Zuriguel}},
  \bibinfo{journal}{Scientific Reports} \textbf{\bibinfo{volume}{7}},
  \bibinfo{pages}{1} (\bibinfo{year}{2017}).

\bibitem[{\citenamefont{Schweitzer and Farmer}(2007)}]{schweitzer2007brownian}
\bibinfo{author}{\bibfnamefont{F.}~\bibnamefont{Schweitzer}} \bibnamefont{and}
  \bibinfo{author}{\bibfnamefont{J.}~\bibnamefont{Farmer}},
  \emph{\bibinfo{title}{Brownian Agents and Active Particles}}
  (\bibinfo{publisher}{Springer}, \bibinfo{year}{2007}).

\bibitem[{\citenamefont{Martinez-Gil et~al.}(2011)\citenamefont{Martinez-Gil,
  Lozano, and Fernández}}]{imartinez2011multi}
\bibinfo{author}{\bibfnamefont{F.}~\bibnamefont{Martinez-Gil}},
  \bibinfo{author}{\bibfnamefont{M.}~\bibnamefont{Lozano}}, \bibnamefont{and}
  \bibinfo{author}{\bibfnamefont{F.}~\bibnamefont{Fernández}}, in
  \emph{\bibinfo{booktitle}{International Workshop on Adaptive and Learning
  Agents}} (\bibinfo{publisher}{Springer}, \bibinfo{address}{Berlin,
  Heidelberg}, \bibinfo{year}{2011}), pp. \bibinfo{pages}{54--69}.

\bibitem[{\citenamefont{Fiorini and Shiller}(1998)}]{fiorini1998motion}
\bibinfo{author}{\bibfnamefont{P.}~\bibnamefont{Fiorini}} \bibnamefont{and}
  \bibinfo{author}{\bibfnamefont{Z.}~\bibnamefont{Shiller}},
  \bibinfo{journal}{The International Journal of Robotics Research}
  \textbf{\bibinfo{volume}{17}}, \bibinfo{pages}{760} (\bibinfo{year}{1998}).

\bibitem[{\citenamefont{Ulicny and Thalmann}(2002)}]{ulicny2002towards}
\bibinfo{author}{\bibfnamefont{B.}~\bibnamefont{Ulicny}} \bibnamefont{and}
  \bibinfo{author}{\bibfnamefont{D.}~\bibnamefont{Thalmann}}, in
  \emph{\bibinfo{booktitle}{Computer Graphics Forum}}
  (\bibinfo{organization}{Wiley Online Library}, \bibinfo{year}{2002}),
  vol.~\bibinfo{volume}{21}, pp. \bibinfo{pages}{767--775}.

\end{thebibliography}
\end{document}